\documentclass[11pt]{article}
\usepackage{amsmath,amssymb,epsfig}

\newcommand{\be}{\begin{equation}}
\newcommand{\ee}{\end{equation}}

\newcounter{unnumber}

\newtheorem{thrm}{Theorem}

\newtheorem{cor}{Corollary}

\newtheorem{prf}[unnumber]{Proof}

\begin{document}

\begin{center}
  {\bf \LARGE Classes of elementary function solutions to the CEV model. I.
}
\end{center}

\vspace{1.3cm} \centerline{
Evangelos Melas $^{a,}$ \footnote{emelas@econ.uoa.gr
}
} \vspace{.6cm}

{\em \centerline{$^a$ Department of Economics, Unit of Mathematics and Informatics,}
\centerline{ Athens, Greece}}

\author{Evangelos Melas}

\begin{abstract}
\noindent


In the equity markets the stock price volatility increases as the
stock price declines.
The classical  Black$-$Scholes$-$Merton (BSM) option pricing model does not reconcile with
this association. Cox introduced the constant elasticity of variance (CEV) model in 1975,
in order to capture this inverse relationship
between the stock price and its volatility.
An important parameter in the model is the
parameter $\beta$, the elasticity of volatility. The CEV model subsumes some of the previous option pricing models. For $\beta=0$,
$\beta=-1/2$, and $\beta=-1$ the CEV model reduces respectively to the BSM model,
the square$-$root model of Cox and Ross,
and the Bachelier model.
Both in the case of the BSM
model and in the case of
the CEV model
it has become  traditional to begin a discussion of option pricing by
starting with the vanilla European calls
and puts.
The pricing formulas for these financial
instruments give concrete information
about the pricing of options only after
the employment of some intermediate
approximation scheme.
However, there are simpler solutions
to both models than those pertaining to
the standard calls and puts.
Mathematically, it makes sense to investigate the simpler cases first.
Furthermore we do
not allow ourselves to be drawn into any rash generalizations or inferences
from the vanilla European case by prematurely focusing on those cases and
we obtain
concrete information for the pricing of options
without needing to introduce
any intermediate approximation schemes.
In the case of BSM model simpler solutions
are the log
and power solutions. These contracts,
despite the simplicity of their mathematical description, are  attracting increasing attention as a trading instrument.
Similar simple solutions have not been studied so far in a
systematic fashion for the CEV model.
We  use Kovacic's algorithm  to derive,
for all half$-$integer values of $\beta$, all solutions ``in quadratures'' of the CEV ordinary differential equation.
These solutions give rise,
by separation of variables,
to simple
solutions to the CEV partial differential equation.
In particular, when $\beta=...,-\frac{5}{2},-2,-\frac{3}{2},-1,  1, \frac{3}{2}, 2, \frac{5}{2},..., $
we obtain four classes of
of denumerably infinite elementary function solutions, when
$\beta=-\frac{1}{2}$ and
$\beta=\frac{1}{2}$ we obtain two classes of
of denumerably infinite elementary function solutions,
whereas,
when $\beta=0$ we
find two elementary function solutions.
In the derived solutions we have
also dispensed with the unnecessary assumption made in the the BSM model asserting
that the underlying asset pays no dividends during the life of the option.


\end{abstract}

\section{Introduction}
\label{nnbddfserhnbcv} \noindent

Approximately 45 years ago Black and Scholes \cite{BS0,BS}, and
independently, Merton \cite{M0} (see also \cite{M}), by making  a number of crucial assumptions,
developed the most widely
used model (hereinafter referred to as the ``BSM model'') in the pricing of financial options.
The BSM model states that by
constantly adjusting the proportions of stocks and
options in a portfolio, the investor can create a
riskless hedge portfolio, where all market risks are
eliminated.
They were led to a PDE,
the Black$-$Scholes$-$Merton
partial differential equation (PDE)
 (hereinafter referred to as the ``BSM PDE''),
which
governs the price of the option over time.

Black and Scholes, and subsequently Merton, derived
\cite{BS,M0} their pricing formula for European calls and puts, by transforming the BSM PDE into the ``heat
equation". The ``heat
equation" is a well known parabolic PDE, which has been
studied extensively by physicists, and describes the evolution over time of the distribution of heat
in a certain region of space under given initial and
boundary conditions.

The pricing formula derived by Black and Scholes, and subsequently Merton, contains the
Gaussian probability density function (see e.g. \cite{G}), and
therefore, some kind of approximation scheme is needed,
in order to extract any concrete information from this
formula for the pricing of options.

The model also assumes that volatility, a measure of the estimation of the future variability for the asset
underlying the option contract, remains constant over the option's life, which is not the case because volatility fluctuates with the level of supply and demand.

The constant volatility hypothesis in BSM model often
leads to results which are inconsistent with market data. To improve the discrepancy, the Constant Elasticity of Variance model
diffusion process (hereinafter referred to as the ``CEV model'') was proposed in \cite{Cox}
(in \cite{Cox1} various jump processes were incorporated into the model) to model the heteroscedasticity and the leverage effect in returns of common stocks. An important
parameter in the model is  the elasticity of volatility $\beta$ which controls the relationship
between volatility and price.

The pricing formula under the CEV diffusion
for European options was given for $\beta<0$ in \cite{Cox}, and for $\beta>0$  in
\cite{Ema}. In summary, the CEV pricing
formula consists of a pair of infinite summations of gamma density and survivor
functions, and its derivation rests on the risk$-$neutral pricing theory \cite{Cox,Ema}. A breakthrough was made in
\cite{Sch} by  Schroder who expressed the pricing formula, for all values of $\beta$, in terms of the non$-$central Chi$-$squared distribution, which facilitates the computations significantly.
 The CEV model has
been further investigated in \cite{Kou, Gat, Y}.

There exists an extensive literature
with different approximation schemes
(see, for instance,
\cite{Far}, \cite{Pos},
 \cite{Di}, \cite{Bab}, \cite{Ben}, and \cite{Dyr}) for the the efficient evaluation of the complementary non$-$central Chi$-$square
distribution function, and only after
the application of such an approximation
scheme concrete information can be
obtained from the pricing formulas
of Cox \cite{Cox} and Emanuel  and MacBeth \cite{Ema} for the pricing
of European options under CEV diffusion.


\subsection{Simpler solutions}

\indent

Both in the case of the BSM
model and in the case of
the CEV model
it has become rather traditional to begin a discussion of option pricing by
starting with the vanilla European calls
and puts. Other instruments are generally labeled as ``exotic".
However, there are simpler solutions
to both models than those pertaining to
the standard calls and puts.

Mathematically, it makes sense to investigate the simpler cases first.
We can get a feeling for the behaviour of options without needing to introduce
any intermediate approximation schemes.
Rather more importantly, we do not
allow ourselves to be drawn into any rash generalizations or inferences from
the vanilla European case by prematurely
focusing on those cases.

In the case of BSM model simpler
solutions are the log
and power solutions. These contracts,
despite the simplicity of their mathematical description, are certainly
not devoid of financial interest \cite{Neu, Neu1, Bou}. On the contrary such contracts are attracting increasing attention as a trading instrument. 

For instance, we note in passing, that
in 1994, Anthony Neuberger, by using the log solution, introduced \cite{Neu} a
simple, yet a very peculiar product, the Log
contract. This contract would forever change
how we look at volatility and lay the
foundation for the introduction of variance
swaps in the late 1990s and the new volatility
index, VIX by CBOE in 2003.

Though
Neuberger may not have realized it then but
what he had done was to initiate
the process by
which volatility would no longer remain a
mathematical and an abstract concept but
become a tangible
asset.

Similar simple solutions have not been studied so far in a
systematic fashion for the CEV model. This is the task we
undertake in this paper. We find classes  of simpler solutions to the CEV model. In the next subsection we define what we mean by
simpler solutions and how we obtain them.



 \subsection{Contribution, simpler
 solutions to the CEV model}

\indent


By using Kovacic's algorithm \cite{Kova},
we find classes of simpler solutions
to the CEV model, for all
half$-$integer
values of $\beta.$ Simpler solutions here means
solutions expressed in terms of elementary
functions. Most importantly we do not need
any intermediate approximation scheme in
order to obtain concrete information from these
solutions for the pricing of the financial instruments they describe under appropriate
initial and boundary conditions.
This is in contradistinction to the pricing
formulas for European calls and puts,
described in Section  \ref{review},
which give information for the pricing of
options only after the employment of an appropriate approximation scheme.


In particular by using
Kovacic's algorithm \cite{Kova},
we
obtain classes of
Liouvillian solutions to the CEV ODE, and we so derive, with separation of variables, solutions       to the
CEV PDE, in terms of elementary functions, for all half$-$integer
values of $\beta.$

Kovacic's algorithm will find all possible Liouvillian solutions
(i.e., essentially, all solutions in terms of quadratures) of
linear second order homogeneous ODEs with complex rational
function coefficients.
Hereafter, solvable
by quadratures means that we consider
 a differential equation “solved” when we are
left only with evaluating an antiderivative.

The Liouvillian solutions we find to the CEV ODE,
are given, for all half$-$integer
values of $\beta$, in section \ref{classes}, and are essentially products
of polynomials, truncated confluent hypergeometric functions of
the first kind, with  exponential functions and powers of the independent variable. These Liouvillian solutions to the CEV ODE
induce, by separation of variables, solutions
in terms of elementary functions, to the CEV PDE.

These
simpler
solutions to the CEV PDE are amenable to analytic
manipulation and easy to use, and thus, are to be juxtaposed with the solutions to the CEV PDE  given
in terms of power series which are described in section \ref{review} in the first category
of the existing literature on the pricing of options under the CEV model.


This paper is organised as follows:
In section \ref{CEVBSM} the bare essentials
of the CEV model are given and the CEV model
is compared with the BSM model.
In section \ref{review} a very brief review
of the literature on the pricing of options under
the CEV diffusion is given.
In section \ref{remark} a remark is made
on the meaning of the term ``closed$-$form'' solutions
which is used in the literature on the
pricing of options, either with the
BSM model or with the CEV model.
In section \ref{Black}
the BSM PDE, and the associated ODE,
are derived. In section \ref{CEV}  the CEV PDE, and the associated ODE, are derived.
In section \ref{alg}
we explain that Kovacic's 
algorithm is an application of  
Picard$-$Vessiot theory to linear second order homogeneous ODEs and
we trace the origin of 
Picard$-$Vessiot theory 
to the 
Galois theory 
of polynomials. 
In section \ref{application} Kovacic's algorithm is applied to the
CEV ODE.
In section
\ref{classes} classes of Liouvillian solutions to the CEV ODE are obtained, and associated solutions to the CEV PDE are derived,  for all
half$-$integer
values of
$\beta$.
In section \ref{future} we outline prospects for future
research. Finally, in the Appendix 
we give 
an outline of Kovacic's algorithm.

\section{ CEV  and CEV versus BSM}

\label{CEVBSM}

The ability to build  a portfolio
of the kind envisaged by Black, Scholes and Merton \cite{BS0,BS,M0} relies on the assumptions of continuous trading and
continuous sample paths of the asset price. Other
assumptions made
in the derivation of the BSM model are:
\begin{itemize}
\item{The asset price follows a lognormal random walk.}

\item{The  volatility of the underlying
asset is constant over the life  of the option contract.}

\item{The risk$-$free interest rate r is constant and known.}

\item{The underlying asset pays no dividends during the life of the
option.}

\item{There are no arbitrage opportunities.}


\item{Short selling is allowed (full use of proceeds from the sale is
permitted).}

\item{Fractional shares of the underlying asset may be traded.}
\end{itemize}

 The BSM model also applies to any
financial instrument the future of which is uncertain
at the present time. Basically it has to do with the
pricing of options, but anything vaguely connected
such as corporate debt is equally grist for its mill.

Volatility, a measure of
how much a stock can be expected to move in the
near$-$term, is probably the most important single
input to any option pricing model. In BSM model it is assumed that  volatility is a constant over time. This means that
the variance of the return is constant over the life
of the option contract and is known to market
participants.

\subsection{The CEV model}

\label{SEV}

\indent

While volatility can be relatively constant in
very short term, it is never constant in longer term. Therefore, a remedy for this shortcoming
of Black-Scholes$-$Merton model  is needed.
The CEV model
introduced by Cox \cite{Cox} is an
example of such a remedy.

The CEV spot price
model is a one$-$dimensional diffusion model with the instantaneous volatility specified to be a power function of
the underlying spot price, $ \sigma(S) = a S^{\beta}.$ It was introduced by Cox
as one of the early alternatives
to the geometric Brownian motion,
assumed in the BSM model,
to model asset prices (see also \cite{Cox1} where
alternative stochastic processes for the asset price
are introduced).

The CEV stochastic process is closely related to Bessel processes
and is analytically tractable.
In fact there are  analytic forms of
option pricing formulas for the CEV diffusion;
for $\beta<0$ was given by  Cox \cite{Cox} and  for  $\beta>0$ was given by   MacBeth and  Merville \cite{Ema}. Subsequently Schroder \cite{Sch}
expressed in both cases the option pricing formulas in terms of the complementary non$-$central Chi$-$square distribution function.




For $\beta = 0$ the CEV model reduces to the constant volatility geometric Brownian motion process employed in the Black, Scholes and Merton
model. When $\beta = -1$, the volatility specification is that
of Bachelier (the asset price has the constant diffusion coefficient, while the logarithm of the asset price has the
$a/S$ volatility).
We mention in  passing that Bachelier's \cite{Bac} work had fallen into oblivion, at least in the financial circles, for an extensive period, and  was rediscovered by mathematical economists such as
Paul Samuelson in the 1960s. Recently it has become again popular since it
assumes that interest rates can be negative.
For $\beta = -1/2$ the model reduces to the
square$-$root model of Cox and Ross \cite{Cox1}.

Thus Cox rather than assuming constant volatility,
expressed the volatility as  a function of the
price of the underlying  asset. This is precisely
the main advantage of the CEV model: The volatility of the underlying asset is linked to its
price level, thus exhibiting an implied volatility smile (or implied volatility skew) that is
a convex and monotonically decreasing function of exercise price, similar to the volatility
smile curves observed in practice (see, for example, \cite{Den}).
The
CEV framework is also consistent with the so$-$called leverage effect (i.e., the existence of
a negative correlation between stock returns and realized stock volatility) as documented
for instance in \cite{Wu}.

\subsection{CEV  versus BSM }

\label{CEVAGBSM}

\indent

In \cite{Mac},
MacBeth and Merville, compared the performances of the
BSM and CEV models through
simulations and real examples. Their results show
that the CEV model
has a better performance, which underprices
in$-$the$-$money call options and overprices out$-$of$-$the money call options.

In \cite{Dia}  Dias and  Nunes showed that a firm that uses the standard geometric Brownian
motion assumption is exposed to significant errors of analysis which may lead to
non$-$optimal investment and disinvestment decisions.
Given that the lognormal assumption with constant volatility does not capture the implied volatility smile effect observed across a wide range of markets and underlying assets Dias and  Nunes used instead
the CEV diffusion process and gave analytical solutions for perpetual American$-$style call and put options under the CEV diffusion.
Their results strongly
highlight the case for moving beyond the simplistic real options models based on the
lognormal assumption to more realistic models incorporating volatility smile effects.

\section{Review of the literature}

\label{review}

\indent

The CEV diffusion process has been extensively
used to obtain the solutions of several flnancial option pricing problems. In particular, the CEV call option pricing formula for valuing European options has been
initially expressed in terms of the standard complementary gamma distribution function by
Cox \cite{Cox} for $\beta < 0$, and by
Emanuel and MacBeth \cite{Ema} for $\beta > 0$.
Schroder \cite{Sch} has subsequently extended the
CEV model by expressing the corresponding formula
in terms of the complementary non$-$central Chi$-$square
distribution function.


There exists an extensive literature devoted to the efficient computation of the complementary non$-$central Chi$-$square
distribution function, with several alternative representations available (see, for instance, \cite{Sch}, \cite{Far}, \cite{Pos},
 \cite{Di}, \cite{Bab}, \cite{Ben}, and \cite{Dyr}).


Things are more complicated
in the case of exotic options such as lookback and barrier options.
Davydov and Linetsky \cite{Dav} evaluated European lookback options
under the CEV process with a model based on the numerical inversion of
the Laplace transform of the option price.
To evaluate
lookback options Linetsky \cite{Lin} used spectral theory.
In \cite{Davy},
Davydov and Linetsky,
priced single$-$barrier and double$-$barrier options under the CEV diffusion process,
using again spectral theory.

In \cite{Cos},  Costabile,
used the binomial process  to approximate the CEV process and priced lookback options.
Boyle, Tian and Imai \cite{Boy0}
tackled the same problem
using Monte Carlo simulations.
In \cite{Boy}, Boyle and Tian,
priced single$-$barrier and double$-$barrier options under the CEV diffusion process in the numerical trinomial lattice framework.


Therefore  the work which has been done so far on the pricing of options
under the CEV diffusion can be conveniently divided into two
categories:

\begin{enumerate}

\item{

In the first category the pricing
formula is given in
closed$-$form \cite{Cox,Ema,Dav},
which,
after Schroder's work \cite{Sch},
is expressed
in terms of  the complementary non$-$central Chi$-$square
distribution function. There exists an extensive literature ( \cite{Far}, \cite{Pos},
\cite{Di}, \cite{Bab}, \cite{Ben}, and \cite{Dyr}) devoted to the efficient
computation of the complementary non$-$central Chi$-$square
distribution function, with several alternative representations and associated approximation schemes  available.
Pricing (exotic) options by
employing spectral theory (see, for instance, \cite{Lin} and  \cite{Davy}) also employs functions represented
by power series, and  falls also into this category.}

\item{In the second category the pricing of (exotic) options
is  attained by discrete approximations of the CEV process:
By a discrete approximation
of the CEV process using the binomial tree
method \cite{Cos}, the trinomial tree method
\cite{Boy}, the Monte Carlo method \cite{Boy0}.}

\end{enumerate}

Needless to say the above review of the literature is
extremely far from being exhaustive and it only highlights two main trends in the current research for pricing options under the CEV diffusion.

\section{On the notion of ``closed$-$form'' solutions}

\label{remark}

\indent

It is appropriate at this point to make a
remark regarding the meaning of
the term ``closed$-$form'' solutions
which is used in the literature on the
pricing of options, either with the
BSM model or with the CEV model.

This term does \it not \normalfont imply, as one might anticipate, that
the solutions at hand give
answers for the
pricing of options without the implementation of
some kind of approximation scheme. On the contrary, as we explain below, both in the case
of the BSM model and in  the case of the CEV model we have to resort to some
kind of approximation in order to value the options by using the pricing formulas.

Black and Scholes found \cite{BS} the celebrated formulas
for the pricing of the vanilla European calls and
puts by transforming the BSM
PDE into the ``heat
equation".

This PDE, which characterises the propagation of heat in a continuous medium, has been extensively studied
in physics. Its fundamental solution, subject to
appropriate boundary conditions, is the Gaussian
density function (e.g. \cite{Paul}, p. 81). Black and Scholes found the formulas for the pricing of options by transforming this solution back to the  initial dependent and independent variables appearing in the  BSM
PDE.

The formulas so derived by Black and Scholes for
the pricing of options are characterised  in the literature as ``closed$-$form'' solutions. However, we have to bear in mind, that in order
to extract any concrete information from them,
we need to compute the integrals $N(d_{i})=\frac{1}{\sqrt{2 \pi}}  \int_{- \infty}^{d_{i}}    e^{- \frac{1}{2} s^{2}} ds,$
where $d_{i}$ are appropriate limits,
numerically by
quadrature rules such as the Simson's rule or
Gaussian rule.

Moreover, in the case of the CEV model the pricing formulas
which have been derived by Cox \cite{Cox} for $\beta < 0$,  by
Emanuel and MacBeth \cite{Ema} for $\beta > 0$,
and by Schroder \cite{Sch}, give answers for the
pricing of options only after the implementation
of some kind of approximation scheme.
For instance, as we noted in subsection
\ref{review}, in the case of Schroder's
pricing formulas, which are expressed in terms
of the complementary non$-$central Chi$-$square
distribution function, there is a whole ongoing
literature ( \cite{Far}, \cite{Pos},
\cite{Di}, \cite{Bab}, \cite{Ben}, and \cite{Dyr}, to mention only a very small fraction of it,) of approximations schemes for its efficient and fast evaluation.

These classes of solutions are to be juxtaposed
with the classes of solutions to the CEV model we derive in this paper when $2 \beta$ belongs to
the integers. The classes of solutions we derive
in this paper do not involve any integrals or
functions expressed in terms of series and
we do not need any
intermediate approximation schemes    in order to use these solutions  for the pricing of various
financial instruments. Thus the solutions we derive in this paper complement the solutions of the first
category, described in subsection
\ref{review}, for the pricing of options
under the CEV diffusion.

\section{Derivation of the BSM PDE and BSM ODE}
\label{Black}

\indent

Black,  Scholes and Merton   made the fundamental
observation \cite{BS0,BS,M0} that if one could \bf{perfectly hedge} \normalfont an option, then one could price it
as well. The reason being that a perfectly hedged portfolio has no uncertainty,
and hence has a risk$-$free rate of return given by the spot interest rate r.

In summary, their derivation
goes as
follows:
Consider a portfolio $\Pi$ which contains one option and $- \Delta$ units of the
underlying asset. The value of the portfolio is
\be
\Pi=V - \Delta \cdot S_{t},
\ee
where $\Delta$ is to be determined,  $S_{t}$ is the asset value
at time $t$, and $V=V(S,t)$ is the value of the
option.

It is assumed that the asset
value has lognormal
dynamics, i.e., it satisfies the SDE

\be
\label{sde}
dS_{t}= r S_{t} dt + \sigma S_{t} dW_{t},
\ee
where $dS_{t}=S_{t+dt}-S_{t}$  is the change in asset value from $t$  to $t+dt$, $ \sigma$ is the standard deviation per unit time (volatility) of the underlying asset value,
and $W_{t}$ is a Wiener process.
Equation (\ref{sde}) says that the percentage
change in asset value from $t$ to $t+dt$ is normally distributed with mean $\mu dt$ and variance $\sigma^{2} dt.$

We assume that $V \in C^{2,1}(R \times [0,T]),$
so by applying Ito Lemma we have that the change in the value of
the portfolio is given by:
\begin{eqnarray}
d\Pi &=&  dV - \Delta \cdot dS_{t} \nonumber \\
&=& \frac{\partial V}{\partial t} dt +
\frac{\partial V}{\partial S} dS_{t} +
\frac{\partial^{2} V}{\partial S^{2}} (dS_{t})^{2}
- \Delta \cdot dS_{t} \nonumber \\
&=& \left ( \frac{\partial V}{\partial t} + \frac{1}{2}
\sigma^{2} S^{2} \frac{\partial^{2} V}{\partial S^{2}}
 \right ) dt + \left ( \frac{\partial V}{\partial S}     - \Delta \right ) dS_{t}.
\end{eqnarray}


If we choose $\Delta = \frac{\partial V}{\partial S}$
 then the change $ d\Pi $ in the value of the portfolio $ \Pi $ is no longer sensitive to random changes in the value $ S_{t} $ of the underlying asset,  and we obtain:
\be
\label{port}
d\Pi = \left ( \frac{\partial V}{\partial t} + \frac{1}{2}
\sigma^{2} S^{2} \frac{\partial^{2} V}{\partial S^{2}}
 \right ) dt.
\ee
Thus the choice $\Delta = \frac{\partial V}{\partial S}$ yields
a perfectly hedged portfolio, i.e., a portfolio
with no
 uncertainty.

Since there is no uncertainty left in the portfolio, its value,
by the principle of no arbitrage,  has to be the same as if being on a bank account with the risk$-$free
interest rate $r.$ If the return of $\Pi$ were larger
than $r$  then one would simply take a loan at a risk$-$free rate $r$ and buy the portfolio $\Pi$
to obtain a guaranteed profit.
Conversely, If the return of $\Pi$ were smaller
than $r$ then one would simply short the
portfolio and invest it in the bank.

However  this is ruled out by the principle of no arbitrage  and therefore we must have:
\begin{eqnarray}
d\Pi & = & r \Pi dt \nonumber \\
\label{port1}
& = & r (V - \Delta \cdot  S_{t})  dt.
\end{eqnarray}
By combining Equations (\ref{port}) and (\ref{port1}) we arrive at
\be
\label{PDE}
\frac{\partial V}{\partial t} + \frac{1}{2}
\sigma^{2} S^{2} \frac{\partial^{2} V}{\partial S^{2}} + r S \frac{\partial V}{\partial S} - r V =0.
\ee
This is the BSM PDE.

For the BSM PDE, which is a backward parabolic equation, we must
specify final and boundary conditions, for
otherwise the PDE does not have a unique solution. For instance for a vanilla European call $c(S,t)$ with exercise price $E$ and expiry date $T$, the final condition is just its payoff at $T$

\be
c(S,T)=max(S-E,0), \  \rm{for \ all} \ \normalfont \it S \geq \rm 0.
\ee
The asset$-$price boundary conditions
are applied at zero asset price, $S=0,$
and as $S \rightarrow +\infty.$ At $S=0$
we have
\be
c(0,t)=0, \  \rm{for \ all} \ \normalfont \it t \geq \rm 0.
\ee
The second boundary condition, as $S \rightarrow +\infty,$ reads
\be
c(S,t) \sim S - E e^{-r (T-t)}, \ as \ S \rightarrow +\infty, \  \rm{for \ all} \ \normalfont \it t \geq \rm 0.
\ee

Moreover, for a vanilla European put option $p(S,t),$ the final condition is the
payoff at $T$
\be
p(S,T)=max(E-S,0), \  \rm{for \ all} \ \normalfont \it S \geq \rm 0.
\ee
The asset$-$price boundary conditions
are applied again at zero asset price, $S=0,$
and as $S \rightarrow +\infty.$ At $S=0$,
assuming that interest rates are constant, we have
\be
p(0,t)=E e^{- r t}, \  \rm{for \ all} \ \normalfont \it t \geq \rm 0.
\ee
As $S\rightarrow +\infty,$ the option is
unlikely to be exercised and so for $t>0,$
we have
\be
p(S,t) \rightarrow 0, \ as \ S \rightarrow +\infty, \  \rm{for \ all} \ \normalfont \it t \geq \rm 0.
\ee

With these final conditions and asset$-$price
boundary conditions Black and Scholes \cite{BS0,BS}, and
independently, Merton \cite{M0} derived their
pricing formulas for European calls and puts.
Their pricing  formulas are essentially the
so called fundamental solution  of the heat
equation (see e.g. \cite{Paul}, p. 81), which is the Gaussian
density function.


We note in passing that the mathematical analysis of American options is more complicated
than that of European options. It is almost always impossible to find a useful explicit
solution to any given free boundary problem, and so a primary aim is to construct
efficient and robust numerical methods for their computation.


In contrast to ODEs
there is no unified theory of PDEs.
Some equations have their own theories,
while others have no theory at all.
The solutions of a PDE form in general
an infinite$-$dimensional space and
there is a wide range of methods
to probe this solution space.

This also applies to the BSM PDE (\ref{PDE}).
This is a second$-$order linear homogeneous
separable backward parabolic PDE.
A method to solve a linear homogeneous
separable PDE is the method of separation
of variables.
In this method such a PDE
involving n independent variables is
converted into n ODEs.

In particular the BSM PDE
(\ref{PDE}) with the separation  of variables
\be
\label{sep}
V(S,t)=A(S)B(t)
\ee
is reduced to the following pair of
ODEs
\begin{eqnarray}
\label{ODE}
\frac{1}{2}
\sigma^{2} S^{2} \frac{\rm{d}^{2} \it{ A}}{\rm{d} \it{S}^{\rm{2}}} + r S \frac{\rm{d} \it{ A}}{ \rm{d} \it{S}} - (r+\lambda) A &=&0, \\
\label{ODE1}
\frac{\rm{d} \it{ B}}{ \rm{d} \it{t}} + \lambda B &=&0,
\end{eqnarray}
where $\lambda$ is an arbitrary constant. We name
 ODE (\ref{ODE}) the BSM ODE. The general solutions to the ODEs (\ref{ODE}) and (\ref{ODE1})
 are given respectively by

\begin{eqnarray}
\label{coef}
 A(S) & = & \alpha S^{\frac{\xi+\zeta}{2 \sigma^{2}}} + \beta S^{\frac{\xi-\zeta}{2 \sigma^{2}}}, \\
 \label{coef2}
 &  &  \rm{where}, \normalfont
 \ \ \xi= \sigma^{\rm{2}} - \rm{2} \it{r}, \ \ \zeta=\sqrt{(\sigma^{\rm{2}} - \rm{2} \it{r})^{\rm{2}} + \rm{8} \sigma^{2}(\it{r}+\lambda)}, \\
\rm{and}, \nonumber \\
\label{coef1}
B(t)&=&\gamma e^{- \lambda t},
\end{eqnarray}
where $\alpha, \beta, \ \rm{and} \ \gamma$ are arbitrary constants.

From Equations (\ref{sep}), (\ref{coef}), and
(\ref{coef1}) the following solution to the
BSM PDE (Equation (\ref{PDE}))  is readily obtained
\be
V(S,t)=e^{- \lambda t} (a S^{\frac{\xi+\zeta}{2 \sigma^{2}}} + b S^{\frac{\xi-\zeta}{2 \sigma^{2}}}),
\ee
where $a, b$ arbitrary constants and $\xi, \zeta$ are specified in Equation (\ref{coef2}).


It is noteworthy that both the fundamental
solution of the heat equation which gives
rise to the pricing formulas of Black, Scholes
and Merton and the solutions to the CEV model we derive in this paper are obtained
by exploiting symmetries, albeit symmetries
of different objects; we elaborate more on this issue
in section \ref{alg}.

The fundamental
solution of the heat equation is
one of the most fundamental formulas
in all of mathematics. It is the famous
Gaussian or normal density of probability
theory. It describes an  initial point source
of heat at a given position.

It
can be derived in a number of different 
ways. Namely,  it can be derived by Fourier transform ( see e.g. \cite{Shaw} p.98, p.99)
and it can also be derived by using the Lie
symmetries of the heat equation. In fact
by using
the Lie symmetries of the heat equation
the fundamental solution
can be derived in two distinct ways.

One derivation results from the invariance
of the heat equation under the one$-$parameter
Lie group of dilations
( \cite{Paul} p.91, p.92 and \cite{Olver} p.119). Another derivation (\cite{Olver}  p.119, p.120) results from a symmetry of the heat equation which cannot be anticipated from basic physical principles and can only be obtained  by applying  Lie theory of symmetries to the
heat equation. Remarkably the last derivation
is the simplest of the three.

The simpler log and power solutions of
the BSM PDE, which not only are they not
devoid of financial interest but also they
are attracting increasing attention as a
trading instrument,
are derived  with  separation of
variables (see e.g. \cite{Shaw},     p. 127, p. 132).
No resort to symmetry principles is needed.

In order to derive classes of  simpler solutions to the CEV model in this paper we follow an hybrid method:
We apply symmetry principles not to the CEV PDE itself,
but to the CEV ODE, which is derived from the CEV PDE
by separation of variables and so we obtain classes of
simpler solutions to the CEV ODE first. Then the classes
of simpler solutions
to the CEV ODE easily
induce classes of simpler solutions to the CEV PDE.

The derivation of the CEV ODE and the CEV PDE is similar to the derivation of the BSM ODE and the BSM PDE, and it is outlined in Subsection \ref{CEVPDE}.



\section{Derivation of the CEV PDE and CEV ODE}

\label{CEV}


\indent

Cox \cite{Cox} considered the CEV diffusion
process governed by the SDE when $q=0$
\be
\label{cevsde}
dS_{t}= (r-q) S_{t} dt + \alpha S_{t}^{\beta+1} dW_{t},
\ee
where $q$ is a dividend yield parameter
and the instantaneous volatility is
specified to be a power function of the
underlying spot price, $\sigma(S)=\alpha S^{\beta}$, $\alpha$  is the volatility
scale parameter. An easy calculation shows that $\beta$ is the elasticity of the volatility $\sigma(S)$.

The elasticity parameter $\beta$ is the
central feature of the model, and controls
the relationship between the volatility
and price of the underlying asset.
The dividend yield parameter $q$ has been
introduced to dispense with the
unnecessary
assumption made in the BSM model asserting
that the underlying asset pays no dividends during the life of the option.
$dS_{t}, r, W_{t}$ remain as defined in
Section \ref{Black}.

The CEV diffusion model seems to be a natural extension of the BSM model. In fact as it is
pointed out in Subsection \ref{CEV}
the CEV model subsumes some of the previous option pricing models. For $\beta=0$,
$\beta=-1/2$, and $\beta=-1$ the CEV model reduces respectively to the BSM model \cite{BS0,BS,M0},
the square$-$root model \cite{Cox1} of Cox and Ross,
and the Bachelier model \cite{Bac}.

In Subsection \ref{CEVAGBSM} is made clear that
the CEV model performs better than the BSM model \cite{Mac,Dia}. Moreover,  in \cite{Beck}, Beckers considered the CEV model and its implications for option pricing on the basis of empirical studies and concluded that
the CEV class could be a better descriptor of the actual stock price
than the traditionally used lognormal model.


\subsection{CEV PDE and CEV ODE}

\label{CEVPDE}

\indent

The derivation of the CEV ODE and the CEV PDE is similar to the
derivation of the BSM ODE and the BSM PDE.
We give it in the detail needed in order to
clarify the similarities and the dissimilarities between the two derivations.


Consider a portfolio $\Pi$ which contains one option and $- \Delta$ units of the
underlying asset. The value of the portfolio is
\be
\Pi=V - \Delta \cdot S_{t},
\ee
where $\Delta$ is to be determined,  $S_{t}$ is the asset value
at time $t$, and $V=V(S,t)$ is the value of the
option.
It is assumed that the asset
value
satisfies the SDE (Equation  (\ref{cevsde}))
$$
dS_{t}= (r-q) S_{t} dt + \alpha S_{t}^{\beta+1}.
$$

We assume that $V \in C^{2,1}(R \times [0,T]),$
so by applying Ito Lemma we have that the change in
the value of the portfolio is given by:
\begin{eqnarray}
d\Pi &=&  dV - \Delta \cdot dS_{t} \nonumber \\
&=& \frac{\partial V}{\partial t} dt +
\frac{\partial V}{\partial S} dS_{t} +
\frac{\partial^{2} V}{\partial S^{2}} (dS_{t})^{2}
- \Delta \cdot dS_{t} \nonumber \\
&=& \left ( \frac{\partial V}{\partial t} + \frac{1}{2}
\alpha^{2} S^{2 \beta + 2} \frac{\partial^{2} V}{\partial S^{2}}
 \right ) dt + \left ( \frac{\partial V}{\partial S}     - \Delta \right ) dS_{t}.
\end{eqnarray}

The choice $\Delta = \frac{\partial V}{\partial S}$ yields
a perfectly hedged portfolio, i.e., a portfolio
with no
 uncertainty. With this choice
 the change
 $ d\Pi $ in the value of the portfolio $ \Pi $
is given by:
\be
\label{portvalue}
d\Pi = \left ( \frac{\partial V}{\partial t} + \frac{1}{2}
\alpha^{2} S^{2 \beta + 2} \frac{\partial^{2} V}{\partial S^{2}}
 \right ) dt.
\ee



Since there is no uncertainty left in the portfolio,
the change $ d\Pi $ in its value $ \Pi $
has to be equal to
\begin{eqnarray}
d\Pi & = & (r-q) \Pi dt \nonumber \\
\label{portvalue1}
& = & (r-q) (V - \Delta \cdot  S_{t})  dt,
\end{eqnarray}
where $r$  is the risk$-$free
interest rate and $q$ is a dividend yield parameter.

By combining Equations (\ref{portvalue}) and (\ref{portvalue1}) we arrive at
\be
\label{CEVPAR}
\frac{\partial V}{\partial t} + \frac{1}{2}
\alpha^{2 } S^{2 \beta + 2} \frac{\partial^{2} V}{\partial S^{2}} + (r-q) S \frac{\partial V}{\partial S} - (r-q) V =0.
\ee
This is the CEV PDE.

For the CEV PDE, which is a backward parabolic equation, we must
specify final and boundary conditions, for
otherwise the PDE does not have a unique solution. For instance for  vanilla European calls and puts
 with exercise price $E$ and expiry date $T$
the final and boundary conditions are those
specified in Section \ref{Black}.

A method to solve a linear homogeneous
separable PDE is the method of separation
of variables.
In particular the CEV PDE
(\ref{CEVPAR}) with the separation  of variables
\be
\label{sepcev}
V(S,t)=C(S)D(t)
\ee
is reduced to the following pair of
ODEs
\begin{eqnarray}
\label{ODECEV}
\frac{1}{2}
\alpha^{2} S^{2 \beta + 2} \frac{\rm{d}^{2} \it{ C}}{\rm{d} \it{S}^{\rm{2}}} + (r-q) S \frac{\rm{d} \it{ C}}{ \rm{d} \it{S}} - (r-q+\lambda) C &=&0, \\
\label{ODECEV1}
\frac{\rm{d} \it{ D}}{ \rm{d} \it{t}} + \lambda D&=&0,
\end{eqnarray}
where $\lambda$ is an arbitrary constant. We name
 ODE (\ref{ODECEV}) the CEV ODE.

 The general solution to the ODE (\ref{ODECEV1})
 is given by
\be
\label{odecev}
D(t)=\delta e^{- \lambda t},
\ee
where $\delta$ is an arbitrary constant. The general solution to the ODE (\ref{ODECEV}) is given for
example in \cite{Dav} and it reads
\be
\label{pdecev}
C(S)= \varepsilon C_{1}(S) + \zeta C_{2}(S),
\ee
where $\varepsilon, \zeta$ are arbitrary constants and $C_{1}(S),$  $C_{2}(S),$
are given respectively by
\begin{eqnarray}
C_{1}(S)=\left \{
  \begin{tabular}{cc}
  $S^{\beta + \frac{1}{2}} \ e^{\frac{\epsilon}{2} x} M_{k,m}(x)$, &  $ \beta<0 $ \\
  \label{c1}
     &  \\
$S^{\beta + \frac{1}{2}} \ e^{\frac{\epsilon}{2} x} W_{k,m}(x)$, &  $\beta>0$
  \end{tabular}
\right .
\end{eqnarray}
and,
\begin{eqnarray}
C_{2}(S)=\left \{
  \begin{tabular}{cc}
  $S^{\beta + \frac{1}{2}} \ e^{\frac{\epsilon}{2} x} W_{k,m}(x)$, &  $\beta<0$ \\
    \label{c2}
   &  \\
$S^{\beta + \frac{1}{2}} \ e^{\frac{\epsilon}{2} x} M_{k,m}(x)$, &  $\beta>0$
  \end{tabular}
\right .
\end{eqnarray}
where $M_{k,m}(x)$ and $W_{k,m}(x)$ are the Whittaker functions (see e.g. \cite{Handbook},  Chapter 13, p.505), and,
\begin{eqnarray}
x&=&\frac{|r-q|}{\alpha^{2} |\beta|} S^{-2 \beta}, \qquad
\epsilon=sign((r-q)\beta), \\
k&=&\epsilon   \left (\frac{1}{2}+\frac{1}{4|\beta|} \right )- \frac{r - q + \lambda}{2 |(r-q)\beta|}, \qquad m=\frac{1}{4|\beta|}.
\end{eqnarray}

Whittaker's functions $M_{k,m}(x)$ and $W_{k,m}(x)$ are defined in terms of the confluent hypergeometric functions of the first
and the second kind $F$ and $U$ respectively  (see e.g. \cite{Handbook},  Chapter 13, p.504$-$505) as follows
\begin{eqnarray}
M_{k,m}(x)&=& e^{-\frac{x}{2}} x^{m +\frac{1}{2} } F\left(m-k+\frac{1}{2}, 1 + 2 m ; x\right), \\
W_{k,m}(x)&=& e^{-\frac{x}{2}} x^{m +\frac{1}{2} } U\left(m-k+\frac{1}{2}, 1 + 2 m ; x\right).
\end{eqnarray}

From Equations (\ref{sepcev}), (\ref{odecev}), (\ref{pdecev}), (\ref{c1})and
(\ref{c2}) the following solution to the
CEV PDE (Equation  (\ref{CEVPAR})) is readily obtained
\be
V(S,t)=e^{- \lambda t} (c C_{1}(S) + d C_{2}(S)),
\ee
where $c, d$ are arbitrary constants.

We note that when $\beta=0$ the solution
space of the CEV ODE (Equation (\ref{ODECEV})) is spanned by the Liouvillian solutions
\begin{eqnarray}
&&S^{\frac{\xi_{1}+\zeta_{1}}{2 \sigma^{2}}} \ \ \rm and  \ \ \it S^{\frac{\xi_{\rm 1}-\zeta_{\rm 1}}{\rm 2 \sigma^{\rm 2}}},
\ \rm{where}, \\
\label{coefficients}
 &  &
 \xi_{1}= \sigma^{\rm{2}} - \rm{2} ( \it{r-q}), \ \ \zeta_{\rm 1}=\sqrt{(\sigma^{\rm{2}} - \rm{2} (\it{r-q}))^{\rm{2}} + \rm{8} \sigma^{2}(\it{r-q}+\lambda)}.
\end{eqnarray}
Therefore, in this case the general solution of the CEV ODE is Liouvillian
and it is given by
\be
\label{liou1}
 A_{1}(S)  =  e S^{\frac{\xi_{1}+\zeta_{1}}{2 \sigma^{2}}} + f S^{\frac{\xi_{\rm 1}-\zeta_{\rm 1}}{\rm 2 \sigma^{\rm 2}}},
\ee
where $e, \ f$  are arbitrary constants.
The solution $ A_{1}(S)$, as expected, is nothing but the solution $ A(S)$
(Equation (\ref{coef})) where $r$
has been replaced by $r-q.$

From Equations (\ref{sepcev}), (\ref{odecev}),
 and
(\ref{liou1}), we conclude
that when $\beta=0$ the following solution to the
CEV PDE (Equation  (\ref{CEVPAR})) is readily obtained
\be
\label{liou2}
V(S,t)=e^{- \lambda t} (g S^{\frac{\xi_{1}+\zeta_{1}}{2 \sigma^{2}}} + h S^{\frac{\xi_{\rm 1}-\zeta_{\rm 1}}{\rm 2 \sigma^{\rm 2}}},
\ee
where $g, \ h$  are arbitrary constants.




\section{Kovacic's algorithm}

\label{alg}

\indent

To the best of our knowledge
Kovacic's algorithm has not been applied
so far to problems
appearing in the field of
Economics (the
only reference
where it appears
that the authors
apply Kovacic's algorithm
to problems
appearing in the field of
Economics
is \cite{kovec},
but in this reference the authors
do not use Kovacic's algorithm
to derive new solutions to
economic models but rather they comment
on the possible applications of
Kovacic's algorithm to certain
financial models).
This is to be juxtaposed with
with the numerous applications
of Lie theory of continuous
symmetries to problems
arising in the field of
Economics (see e.g. \cite{LA} and references
therein).



Kovacic's algorithm is the outcome of
Picard$-$Vessiot  theory when this is applied to linear second order homogeneous
ODEs with rational function coefficients. Interestingly enough
Picard$-$Vessiot  theory and Lie theory have sprung from the
same origin, Galois theory of polynomials. In subsection
\ref{his}  we highlight the main results of these theories and
how these are related to the main result of their origin,
Galois theory of polynomials.

By so doing we bring to the fore
one of the major connections
between differential equations and abstract algebra  $-$    one that is not
commonly emphasized. This connection is related to the solvability of
the equation. In the case of abstract algebra, we have the result that
the polynomial equation is solvable by radicals if and only if the Galois
group of the polynomial is a solvable group. The differential equation
result is that if the equation admits a solvable group, then it is solvable
by quadratures.



\subsection{Galois Theory, Lie Theory, Picard$-$Vessiot
theory }

\label{his}

The idea behind the Galois theory of
polynomials \cite{Galois,Galois1,Galois2} is to associate to a polynomial a group, the Galois group, which is  the group of symmetries
of the roots that preserve all the algebraic relations among these
roots,  and deduce properties of the roots from properties of this group.

For instance,  a solvable Galois group implies that the roots can be expressed
in terms of radicals, and  the fact that the
symmetric groups $S_{n}$ for $n\geq 5$  are
not solvable accounts for the fact that there is
no generic formula for extracting the roots
of polynomial equations of degree five or more.

Let $p(x)=0$ be an irreducible polynomial equation with
coefficients in a field $\mathcal F.$
Every element of the Galois group of $p(x)$
permutes the roots of $p(x)$ and
leaves invariant all the algebraic relations
satisfied by the roots.



This allows to describe the Galois group of  $p(x)$  as a group of automorphisms of a field extension:
Let $\mathcal A$ be the group of automorphisms
of the field extension of $\mathcal F$
which is formed from $\mathcal F$ by adjoining to it the roots of the polynomial equation $p(x)=0.$ Galois group is the subgroup of
$\mathcal A$ which leaves fixed pointwise
the elements of $\mathcal F$.




Galois theory motivated two major developments
in the theory of differential equations:
Lie group theory \cite{Lie1,Lie2} for continuous symmetries
of differential equations, and, Picard$-$Vessiot
theory  \cite{Piccard,Vessiot,Vessiot1} of the differential field extensions
generated by the solutions of a linear differential
equation.

Picard$-$Vessiot
theory is closer in spirit to Galois theory
than Lie theory. In Picard$-$Vessiot
theory one replaces fields by differential fields:
fields with a derivation $D$. Just as adjoining a root of a polynomial equation to
a field gives an extension of fields, adjoining a root of a differential equation to
a field gives an extension of differential fields.

As in Galois theory, one can form
the differential Galois group
of an irreducible linear homogeneous ordinary
differential equation $\mathcal E$ with coefficients
in a field $\mathcal F(x)$. The differential Galois group of $\mathcal E$ is the group of automorphisms of the extension of the differential field  $\mathcal F(x)$ which leaves the elements of $\mathcal F(x)$
fixed.


A main result of Picard$-$Vessiot
theory  is that a linear homogeneous differential equation can be solved by
quadratures
if and only if its differential Galois group
is solvable. Picard$-$Vessiot
theory falls into the realm of differential
algebra.



Lie's motivation in
constructing his theory was
the desire to extend
the theory developed by Galois
(and Abel \cite{Abel})
to differential equations.
If the discrete invariance group of an algebraic
equation could be exploited to generate algorithms to solve the algebraic
equation by radicals,
might it be possible that the continuous
invariance group of a differential equation could be exploited to solve
the differential equation by quadratures?


In fact Lie  showed in 1894 \cite{Lie2} that as an algebraic equation of degree n is solvable by radicals if its Galois group
is an $n$th$-$order solvable group, an $n$th$-$order ordinary differential equations can be integrated by quadratures if it admits a solvable $n$$-$parameter symmetry group.

Two are the key  ideas which lie at the
foundations of Lie theory and enable its development: The first key idea is Lie's
great advance to replace the complicated,
nonlinear invariance condition of an analytic
 function
under a one$-$parameter Lie group of
 transformations by a vastly more useful
linear infinitesimal condition and to recognize that if an analytic function satisfies the
infinitesimal condition then it also satisfies
the finite condition and vice versa.

The second key idea in Lie theory
is his somewhat unique view of
differential equations:
ODEs and PDEs can be
viewed
as locally analytic functions
in a space whose coordinates are
independent variables, dependent variables, and the various derivatives of one with respect to
the other.

This implies in particular that
we can derive
one$-$parameter Lie groups which leave  ODEs or PDEs invariant by applying
the first key idea.
It turns out (see e.g \cite{Cant}  p. 129) that
this can be achieved in a
two$-$step process.
In the first step we derive
the vector fields whose integral curves
are the orbits of the one$-$parameter Lie groups which leave the ODEs or PDEs invariant.
In the second step we use these vector fields
to derive the one$-$parameter Lie groups
which leave the ODEs or PDEs invariant.




This makes apparent that Lie  theory falls
into the realm of local differential geometry.
Sophus Lie was at heart a geometer,
and it was through this lens that he viewed much of his work. On the other hand
Picard$-$Vessiot theory falls into
the realm of differential algebra.
Lie theory applies to any differential
equation, whereas Picard$-$Vessiot theory
applies only to linear homogeneous ODEs of
$n^{th}$ order.

One expects that
in the case of linear homogeneous ODEs of
$n^{th}$ order the two theories should
be related to each other since
a main result in
both
theories
is that if a linear ODE of
$n^{th}$ order
admits a solvable group, then it is solvable
by quadratures.
However
the links between
Lie theory and Picard$-$Vessiot theory
remained hidden for a long time, mostly because of
the apparent walls that separate the mathematical disciplines of local differential
geometry and differential algebra.

In fact evidence was given \cite{Ou} to
support the common wisdom
that the two theories are not
related to each other.
It came as a surprise when
Ibragimov
found a bridge \cite{Ibr} between Lie symmetries
and Galois groups:
He constructed the Galois groups for
several simple algebraic equations by first calculating their Lie symmetries and
then restricting the symmetry group to the roots of the equation in question.
Thereafter more papers have appeared
\cite{Int} which study the interplay and
connections between the differential
Galois group and Lie symmetries of
linear homogeneous ODEs of
$n^{th}$ order.



We have emphasized that a main result in
Picard$-$Vessiot theory
is that if a linear homogeneous ODE
of
$n^{th}$ order admits a solvable group, then it is solvable
by quadratures. In fact
Picard and Vessiot proceeded
even further \cite{Piccard,Vessiot,Vessiot1} and stated
sufficient and necessary conditions
for the existence of Liouvillian solutions
to a linear homogeneous ODE of
$n^{th}$ order.

Roughly speaking Liouvillian solutions are solutions in quadratures
which can be expressed in terms of exponentials,
integrals and algebraic functions;
a more precise definition of Liouvillian
solutions  is given in subsection
\ref{Liou}. A formal modern proof of
the criteria given by
Picard and Vessiot for the existence of
Liouvillian solutions
was given by Kolchin
\cite{kolchin1,kolchin2}.
These  results are such that  lead to several algorithms \cite{Boul,Singer2} to decide if a linear homogeneous ODE of
$n^{th}$ order has a Liouvillian solution.


Picard$-$Vessiot$-$Kolchin theory and the
ensuing algorithms to decide if a linear homogeneous ODE of
$n^{th}$ order admits  Liouvillian solutions
can be simplified for ${\rm 2}^{nd}$ order
 linear homogeneous ODEs
because of several facts that are summarized in
\cite{Singer} (Chapter 4.3.4).
The resulting algorithm is essentially
the algorithm presented by Kovacic in
\cite{Kova}.
Kovacic's algorithm \cite{Kova} predated and motivated
much of the work on Liouvillian solutions of general linear
homogeneous ODEs of
$n^{th}$ order.
A beautiful account of Picard$-$Vessiot$-$Kolchin theory and
of Kovacic's algorithm is given in Singer's
lectures \cite{Singer1}.

\subsection{Liouvillian Solutions of Second Order Equations}

\label{Liou}

Kovacic's algorithm \cite{Kova} finds a ``closed$-$form'' solution
of the differential equation \be \label{ant} y''+ay'+by=0 \ee
where $a \, {\rm and} \, b$ are rational functions of a complex
variable $x$, provided a ``closed$-$form'' solution exists. The
algorithm is so arranged that if no solution is found, then no
solution can exist. The ``closed$-$form'' solution means a
Liouvillian solution, i.e. one that can be expressed in terms of
algebraic functions, exponentials and indefinite integrals. (As
functions of a complex variable are considered, trigonometric
functions need not be mentioned explicitly, as they can be written
in terms of exponentials. Logarithms are indefinite integrals and
hence are allowed).

In more concrete terms, a Liouvillian function, is a function of one complex
variable, which is the composition of a finite number
of arithmetic operations ($+,-,\times,\div$), exponentials,
constants, solutions of algebraic equations,  and antiderivatives.
It follows directly from the definition   that the set of Liouvillian functions is closed under arithmetic operations, composition, and integration. It is also closed under differentiation. It is not closed under limits and infinite sums.

All elementary functions are Liouvillian. Examples of well$-$knownn functions which are Liouvillian but not elementary
are the nonelementary integrals, for example: The error function,  the Fresnel integrals. \it{All} \normalfont Liouvillian solutions
are solutions of algebraic differential equations, but not conversely.

Examples of functions  which are solutions of algebraic differential equations but not Liouvillian include: the Bessel functions (except special cases), and the hypergeometric functions (except special cases). Such
a special case are the
\it{truncated} \normalfont confluent hypergeometric functions which are going to be of particular importance
in this study. More generally, all functions which are represented as power series (\it{not truncated}\normalfont) \normalfont are \it{not} \normalfont Liouvillian. A more precise definition involves the notion
of Liouvillian field \cite{Kova} and it is not
going to be given here.


Let $\eta$ be a (non$-$zero) Liouvillian solution of the
differential equation (\ref{ant}). It follows that every solution
of this differential equation is Liouvillian. Indeed the method of
reduction of order produces a second solution, namely \be
\label{kritskrits} \eta \int \frac { e^{-\int a}}{\eta ^{2}}. \ee
This second solution is evidently Liouvillian and the two
solutions are linearly independent. Thus any solution, being a
linear combination of these two, is Liouvillian.

A well$-$known
change of dependent variable may be used to eliminate the term
involving $y'$ from the differential equation (\ref {ant}). Let
\be \label{ursuz1} z=e^{\frac {1}{2} \int a} y. \ee
Then Equation (\ref {ant}) yields
\be
\label{ursuz} z''+ \left ( b- \frac{1}{4} a^{2}- \frac {1}{2}a'
\right )z=0. \ee

~Equation (\ref{ursuz}) still
has rational function coefficients and evidently (see Equation
(\ref {ursuz1})) $y$ is Liouvillian if and only if $z$ is
Liouvillian. Thus no generality is lost by assuming that the term
involving $y'$ is missing from the differential Equation
(\ref{ant}). Before giving the main result obtained by Kovacic
\cite{Kova} we first introduce some notation and some terminology.
$C$ denotes the complex numbers and $C(x)$ the rational functions
over $C$.
A  function $\omega$ of $x$
is called an algebraic function
of degree k,
where k is a positive integer, when $\omega$ solves an irreducible
algebraic equation \be \label{doladola} \Pi(\omega,x)=\sum_{{\rm
i}=0}^{{\rm k}} \frac{{\rm P}_{{\rm i}}(x)} {({\rm k}-{\rm i})!}
\omega^{{\rm i}}=0 \ee where ${\rm P}_{{\rm i}}(x)$ are rational
functions of $x$. Let $\nu \in C(x)$ (to avoid triviality, $\nu
\in \!\!\!\!\!| \,\,C$). Then the following holds (\cite{Kova})
\begin{thrm}
Equation \be \label{katrouba} z''=\nu z, \, \, \, ,\, \, \, \,
\nu \in C(x) \ee has a Liouvillian solution if and only if it has
a solution of the form \be z=e^{\int \omega {\rm d}x} \ee where
$\omega$ is an algebraic function of x of degree 1,2,4,6 or 12.
\end{thrm}

The search of Kovacic's algorithm for $\Pi(\omega,x)$ is based on
the knowledge of the poles of $\nu$ and consists in constructing
and testing a finite number of possible candidates for
$\Pi(\omega,x)$. If no $\Pi(\omega,x)$ is found then the
differential Equation (\ref{katrouba}) has no Liouvillian
solutions. If such a
 $\Pi(\omega,x)$ is found and $\omega$ is a solution of the Equation
(\ref{doladola}) then the function $\eta=e^{\int\omega{\rm d}x}$
is a Liouvillian solution of (\ref {katrouba}). If \be
\label{katrouba1} \nu(x)=\frac{s(x)}{t(x)}, \ee
 with $ s,t \in C[x]$,
($C[x]$ denotes the polynomials over C), relatively prime, then
the poles of $\nu$ are the zeros of $t(x)$ and the order of the
pole is the multiplicity of the zero of $t$. The order of $\nu$ at
$\infty$ , ${\rm o}(\infty)$, is defined as \be
\label{oinf}
{\rm
o}(\infty)={\rm max}(0, 4+{\rm d}^{{\rm o}}s- {\rm d}^{{\rm o}}t)
\ee where ${\rm d}^{{\rm o}}s$ and  ${\rm d}^{{\rm o}}t$ denote
the leading degree of $s$ and $t$ respectively (for a
justification of this definition see \cite{Duval} page 10; Kovacic
originally gave a different definition, see \cite{Kova} page 8).
In the Appendix  we give an outline of Kovacic's algorithm. This is the outline of an improved version of the algorithm given by Duval and
Loday$-$Richaud \cite{Duval}.


Now, besides
the original formulation of this algorithm \cite{Kova} we have its several versions and
improvements \cite{Duval}, \cite{Duval1}, \cite{UW} and extensions to higher order equations \cite{SU1}, \cite{SU2}, \cite{HR}. The formulation of the Kovacic algorithm given in \cite{UW} is alternative to its original form \cite{Kova} and
to that presented previously. It seems that it is much more convenient for computer implementation and it has been implemented in Maple.
However, for differential equations with simple structure of singularities, and depending on
parameters it seems that the previous form of the algorithm, which is the form given in \cite{Duval},
is well suited.
Moreover, the original formulation of the algorithm \cite{Kova} consists in fact of three separated algorithms
each of them repeating similar steps. In \cite{Duval} one can find a
modification of the original formulation unifying and improving these three algorithms in one. This
form is very convenient for applications and it is the one employed here.

\section{Application of Kovacic's algorithm}



 \label{application}

\indent

We apply Kovacic's algorithm to the CEV ODE
(Equation (\ref{ODECEV}))
$$
\frac{1}{2}
\alpha^{2} S^{2 \beta + 2} \frac{\rm{d}^{2} \it{ C}}{\rm{d} \it{S}^{\rm{2}}} + (r-q) S \frac{\rm{d} \it{ C}}{ \rm{d} \it{S}} - (r-q+\lambda) C =0
$$
Dividing both sides of Equation (\ref{ODECEV})
by the coefficient of $\frac{\rm{d}^{2} \it{ C}}{\rm{d} \it{S}^{\rm{2}}}$ we  bring it to the form of  Equation (\ref{ant})
\be
\label{redf}
  \frac{\rm{d}^{2} \it{ C}}{\rm{d} \it{S}^{\rm{2}}} + \frac{2 (r-q)}{\alpha^{2} S^{2 \beta + 1} }  \frac{\rm{d} \it{ C}}{ \rm{d} \it{S}} - \frac{2 (r-q+\lambda)}{\alpha^{2} S^{2 \beta + 2}   } C =0.
\ee

A well$-$known
change of dependent variable (Equation (\ref{ursuz1})) may be used to eliminate the term
involving $\frac{\rm{d} \it{ C}}{ \rm{d} \it{S}}$ from the differential equation (\ref {redf}). Let
\be
\label{chvar}
\mathcal C=e^{\frac {1}{2} \int \frac{2 (r-q)}{\alpha^{2} S^{2 \beta + 1} } \rm{d} \it{S} } C.
\ee
Then Equation (\ref{redf}) yields
\be
\label{cevodenew}
\frac{\rm{d}^{2} \mathcal{ C}}{\rm{d} \it{S}^{\rm{2}}} - \nu \mathcal{ C}=0,
\ee
where
\be
\label{nnn}
\nu=\frac{(q-r)^{2} + \alpha^{2}((2 \beta - 1) (q-r) +  2 \lambda
)S^{2\beta}
}{\alpha^{4} S^{2+4\beta}}
\equiv  \frac {s(S)}{t(S)}.
\ee
We note that Equation (\ref{chvar}) implies
that $\mathcal C$ is Liouvillian if and only if
$C$ is Liouvillian. Thus it suffices to apply
Kovacic's algorithm to Equation (\ref{cevodenew}).

Two remarks are now in order regarding Equation
(\ref{cevodenew}):
\begin{enumerate}
\item{Kovacic's algorithm can be applied to Equation (\ref{cevodenew}) if $\nu$
is a ratio of relatively prime polynomials
$s(S)$ and $t(S)$ of
$S$. This implies in particular that $2 \beta$ has to be an integer.}

\item{Kovacic's algorithm starts with assigning orders to the poles of $\nu$,
    i.e. to the zeros of $t(S)$, and the order of the pole is the multiplicity of the zero of  $t(S)$. It also assigns
    order to $\infty$. The order of $\nu$ at
$\infty$, ${\rm o}(\infty)$, is defined as
(Equation (\ref{oinf}))
$ {\rm
o}(\infty)={\rm max}(0, 4+{\rm d}^{{\rm o}}s- {\rm d}^{{\rm o}}t)
$ where ${\rm d}^{{\rm o}}s$ and  ${\rm d}^{{\rm o}}t$ denote
the leading degree of $s$ and $t$. This is related to the order of $\infty$ as a zero of   $\nu$. The cases to be considered in order to
decide if Equation (\ref{cevodenew}) admits
Liouvillian solutions depend crucially on
the orders associated to the poles of $\nu$
and to $\infty$.
}
\end{enumerate}

The first remark restricts $\beta$ to take
half$-$integer values only.
The second remark implies that
if different half$-$integer values of $\beta$
result in different orders of the
poles of $\nu$, and of $\infty$, then
these half$-$integer values of $\beta$ have
to be considered separately.


In the case of Equation (\ref{cevodenew}), $\nu$
has one single pole, namely the number 0. Let
$ {\rm o}(0$) denote the order of 0.
An easy calculation gives

\vspace{0.2cm}

\begin{center}

\begin{tabular}{lccr} 
   & \boldmath   $   {\rm
o}(0$)  & \boldmath   $   {\rm
o}(\infty$)        \\
   \textbf{  2 \boldmath  $  \beta$ = 2,3,...} & 2 + 4$|\beta|$ & 0  \\
   \textbf{ 2 \boldmath $\beta$ = 1} & 4 & 1 \\
   \textbf{ 2 \boldmath $\beta$ = 0} & 2 & 2 \\
   \textbf{ 2 \boldmath $\beta$ = $-$1} & 1 & 4 \\
   \textbf{ 2 \boldmath $\beta$ = $-$2,$-$3,...} & 0 &   2 +  4$|\beta|$   \\
   \end{tabular}

\end{center}

\noindent
The structure of Kovacic's algorithm is such that
the cases $2 \beta=2,3,...$ and $2 \beta=-2,-3,...$ can be considered together.
Thus we are left with five cases we have to consider
\begin{eqnarray}
\bf  1^{st}  \ \ case   && \textbf{  2 \boldmath  $  \beta$ = 2,3,...} \nonumber \\
\bf  2^{nd}  \ \ case   && \textbf{ 2 \boldmath $\beta$ = 1}\nonumber \\
\label{per}
\bf  3^{nd}  \ \ case   && \textbf{ 2 \boldmath $\beta$ = 0}  \\
\bf  4^{th}  \ \ case   && \textbf{ 2 \boldmath $\beta$ = $-$1}\nonumber \\
\bf  5^{th}  \ \ case   && \textbf{ 2 \boldmath $\beta$ = $-$2,$-$3,...}\nonumber
\end{eqnarray}

The implementation of Kovacic's algorithm
in each one of these cases is by no means
trivial and it poses its own problems and challenges. We present here in detail the
application of Kovacic's algorithm in the
first case and report the results in all the
other cases. To give in detail the application
of Kovacic's algorithm in all cases would take
us too far afield. Detailed application of
Kovacic's algorithm in the other four cases will
be given elsewhere \cite{melas}.

\subsection{Application of Kovacic's algorithm when \textbf{  2 \boldmath  $  \beta$ = 2,3,...}}
\label{Appl}

We apply
Kovacic's algorithm to Equation (\ref{cevodenew})
$$
\frac{\rm{d}^{2} \mathcal{ C}}{\rm{d} \it{S}^{\rm{2}}} = \nu \mathcal{ C}.
$$
We note that in the Appendix
 we give only  part of Kovacic's algorithm.
It is the part which is necessary for the application of Kovacic's algorithm to the
case
$2     \beta = 2,3,...  $. For the
complete version of  Kovacic's algorithm
we use in this paper see \cite{Duval}.


\noindent \bf {Input:} \normalfont
From Equation (\ref{nnn}) we have
$$
\nu=\frac{(q-r)^{2} + \alpha^{2}( (2 \beta -1) (q-r)  +  2 \lambda
)S^{2\beta}
}{\alpha^{4} S^{2+4\beta}}
\equiv  \frac {s(S)}{t(S)}.
$$
\noindent The partial fraction expansion for $\nu$ is the
following
\be
\label{poustpoust1} \nu(S)  =
\frac{\left( \frac{q-r}{\alpha^{2}}   \right)^{2}}{S^{2 + 4 \beta}} + \frac{ \frac{
(2 \beta - 1)(q-r) +  2 \lambda
}{\alpha^2}}{S^{2 + 2 \beta}}.
\ee



\noindent
\bf{First step:} \normalfont


\noindent $\textbf{1a.}$ \hspace{0.1cm} ${\rm From \: \; Equation \; (\ref{nnn})
 \;we    \; obtain}$
\begin{eqnarray}
\label{poustpoust} t(S)=S^{2 + 4 \beta}.\;{\rm Hence,}\; m=2 + 4 \beta
&
\Gamma'=\{0\}, & \Gamma=\{0,\infty\}  \\
{\rm o}(0) =2 + 4 \beta& {\rm d}^{\rm o}s = 2 \beta& {\rm d}^{\rm o}t= 2 + 4 \beta \\
\label{order}
{\rm o}(\infty)=
 &
 max(0,4+2 \beta - 2 - 4 \beta )=
 &
max(0,2 - 2 \beta) = 0 \qquad
\\
\label{gamma1}
m^{+}=max(m,{\rm o}(\infty))= &max(2 + 4 \beta,0)=2+4 \beta  & \Gamma_{0}= \{\infty\}  \\
\label{gamma2}
 \Gamma_{2+4 \beta}=\{0 \}.
&
&
\end{eqnarray}


\noindent $\textbf{1b.}$ \hspace{0.5cm} ${\rm Equations  \: \;
(\ref{poustpoust})  \; \;give  }$ \be \label{holoholo}
\gamma_{2}=0 \; \; \; \; \; {\rm and} \; \; \; \; \;
\gamma=\gamma_{2}=0. \ee




\noindent $\textbf{1c.}$ \hspace{0.5cm} ${\rm Equations    \; \;
(\ref{order}) \;  and \;
(\ref{holoholo})  \; \; imply    }$ \be {\rm L}=\{1\}. \ee

\noindent $\textbf{1d.}$ \hspace{0.5cm} \be \label{toratora} {\rm
n}=1. \ee


\noindent
$\!\!\!$ \bf  {Second step:} \normalfont


\noindent {\bfseries 2a.} \hspace{0.5cm} \normalfont \hspace{-0.2cm}
$\infty \in \Gamma_{0}$ and consequently E$_{\infty}=h(\rm n)\{0,1,..., \rm n\}=\it h(\rm1)\{\rm 0, \rm 1\}$. Therefore
\be
\label{Einfty}
\rm E_{\infty}=\{0,1\}.
\ee

\noindent {\bfseries 2b.} \hspace{0.5cm} \normalfont We have n=1 (Equation (\ref{toratora})), $0 \in \Gamma_{2(1 + 2 \beta)}$ (Equation (\ref{gamma2})), ${\rm q}=1+2 \beta > 2,$ since
$2 \beta=2,3,... \ .$ Equation (\ref{MilkyWay}) gives
\be \label{nu1} \left [ \sqrt \nu
\,\right ] _{0}=
\frac{{\rm a}_{0}}{S^{1+2 \beta}}+
\sum_{{\rm i}=2 \beta}^{2} \frac{\mu_{{\rm i},0}}{S^{{\rm i}}}.
\ee
From Equations (\ref{poustpoust1}) and (\ref{nu1})
by using undetermined coefficients we obtain
\begin{eqnarray}
\label{un1}
{\rm a}_{0}^{2} &=& \left (  \frac{q-r}{\alpha^{2}}     \right )^{2}, \\
\label{un2}
\mu_{2 \beta,0} &=& \mu_{2 \beta -1,0}=\mu_{2 \beta -2,0}=...=\mu_{2,0}=0.
\end{eqnarray}
Thus there are two possibilities for
$ \rm a_{0}$,
one being the negative of
the other, and any of these  may be chosen.
We choose
\be
\label{a}
\rm a_{0}=\frac{\it q-r}{\alpha^{2}}.
\ee

Equation (\ref{MilkyWay1}) yields
\be \label{exp} \nu-\left [ \sqrt \nu \,\right ]
_{0}^{2} = \frac{\rm b_{0}}{S^{2+2 \beta}}
+ {\rm O} \left ( \frac{1}{S^{1 + 2 \beta}}      \right ).\ee
Equations  (\ref{poustpoust1}), (\ref{un1}), (\ref{un2})  and (\ref{exp}) imply
\be
\label{b}
\rm b_{0}=\frac{ (2 \beta -1) \it (q-r)
+ \rm 2 \lambda
}{\alpha^2}.
\ee
From Equations (\ref{a})  and (\ref{b}) we have

\noindent
\begin{eqnarray}
{\rm E}_{0}&=&\left \{ \frac
{1}{2} \left ( \rm q +\epsilon \frac{{\rm b}_{0}}{{\rm
a}_{0}} \right ) \left | \right. \ \epsilon= \underline{+} 1 \right \} \nonumber \\
\label{elinf}
&=& \left \{ 2 \beta + \frac{\lambda}{q-r}
 , 1 - \frac{\lambda}{q-r} \right \}.
\end{eqnarray}
From Equation (\ref{blooo}) it follows that   the function ``Sign'' S with domain ${\rm
E}_{0}$ is defined as follows
\be
\label{dom}
{\rm S} \left ( 2 \beta + \frac{\lambda}{q-r}
\right )=1, \quad {\rm S} \left ( 1 - \frac{\lambda }{q-r}
\right )=-1.
\ee

\noindent
\bf  {Third Step:} \normalfont


\noindent {\bfseries 3a.} \hspace{0.5cm} \normalfont For each
family $\underline{{\rm e}}= \left ({\rm e}_{c} \right ) _{c \in
\Gamma}$ of elements ${\rm e}_{c} \in {\rm E}_{c}$ we calculate
the degree ${\rm d} \left ( \underline{{\rm e}} \right )$ of the
corresponding, prospective polynomial P. The sets ${\rm E}_{c}$
are given by Equations
(\ref{Einfty}) and
(\ref{elinf}).



\begin{tabular}{rlllc}
\boldmath ${\rm e}_{0}
\;\;\;
$ &
\quad
\quad
\qquad
&\boldmath${\rm e}_{\infty}\;\;\;
\qquad
$ &\boldmath${\rm d}=1-\displaystyle\sum_{c\in
\Gamma}
{\rm e}_{c}\;\;\;
\qquad
$ & ${\bf Families}$ \\
& & & & \\
\label{first}
$\; \! \! \! \!
1 - \frac{\lambda}{q-r}
$
&
&  \; 0
&
\qquad
$\frac{\lambda}{q-r}$
& F1 \\
& & & & \\
\label{second}
$\;
\; \! \! \! \!
1 - \frac{\lambda}{q-r}
$ &
& \; 1
& $  \;   \; \;   -1 + \frac{\lambda}{q-r}$
& F2 \\
& & & & \\
\label{third}
 $ 2 \beta + \frac{\lambda}{q-r}$ &
& \; 0
& $\! \! \! \! \! \quad 1 - 2 \beta -
\frac{\lambda}{q-r}   $
& F3 \\
& & & & \\
\label{fourth}
$\; \; \; \;
\; \; \; \; \; \; \; \;
\; \; \; \;
2 \beta + \frac{\lambda}{q-r}$ &
& \; 1
& $\! \! \!
\quad - 2 \beta - \frac{\lambda }{q-r}   $
& F4
\end{tabular}
\normalfont

\vspace{0.2cm}
\normalfont \noindent In the last column we
enumerated the different families.


\hspace{0.6 cm}

\noindent {\bfseries 3b.} \hspace{0.6cm}
If d is a non$-$negative integer n, the family should be
retained, otherwise the family is discarded.
This makes $\lambda$, in each of the families
F1, F2, F3, and F4,  to become a
function of $q,r,\beta, \rm and \ \rm n.$

\begin{eqnarray}
\boldmath
{\bf Families}
\qquad
\qquad
\qquad
\qquad
\qquad
\qquad
\qquad
&
& \boldsymbol
 \lambda
\nonumber \\
\label{ygflygflygflty}
{\rm F1}
\qquad \qquad
\quad \qquad \qquad \qquad
\qquad \qquad
&
&
\! \! \! \! \! \! \! \! \! \! \! \!
\! \!
\lambda=\rm n \it (q-r)
\\
&& \nonumber \\
\label{gravity7}
{\rm F2}
\qquad \qquad
\quad \qquad \qquad \qquad
\qquad \qquad
&
&
\! \! \! \! \! \! \! \! \! \! \! \!
\! \!
\! \!
\! \! \! \! \! \! \!
\lambda=(\rm n + 1) \it (q-r)
\\
&& \nonumber
\\
\label{gravity8}
{\rm F3}
\qquad \qquad
\quad \qquad \qquad \qquad
\qquad \qquad
&
&
\! \! \! \! \! \! \! \! \! \! \! \!
\! \! \! \!  \! \! \! \! \! \!
\! \! \! \!
\! \! \! \!
\lambda=-(2 \beta  + \rm n - 1) \it (q-r) \\
&& \nonumber \\
\label{gravity9}
{\rm F3}
\qquad \qquad
\quad \qquad \qquad \qquad
\qquad \qquad
&
&
\! \! \! \! \! \! \! \! \! \! \! \!
\! \! \! \!  \! \! \! \! \! \!
\! \! \!
\lambda=-(2 \beta  + \rm n ) \it (q-r)
\end{eqnarray}



\noindent
{\bfseries 3c}. \hspace{0.7 cm} For each family  retained from
step  \textbf{3b}, we form the rational function $\theta$ given by
Equation (\ref{krookraa}). Since
from Equations  (\ref{gamma1}),
(\ref{gamma2}) and (\ref{toratora}) we have respectively
$\Gamma_{0}=\{  \infty    \}$,
$\Gamma_{2+4 \beta}=\{0 \}$, and n=1,
Equation (\ref{krookraa})
implies \be \label{kliklii} \theta=\frac{{\rm e}_{0}}{S} +
{\rm S}({\rm e}_{0}) \left [ \sqrt
\nu \right ]_{0}, \ee where ${\rm e}_{c}$ denotes any element
of ${\rm E}_{c}$,
${\rm S}({\rm e}_{0})$ are given by
Equations  (\ref{dom}) and $ \left [ \sqrt \nu \right
]_{0}$ is given by Equations
 (\ref{nu1}), (\ref{un1}), and (\ref{un2}).
 By making use of
Equation (\ref{kliklii}) for each of the retained families we
obtain \newline
\begin{eqnarray}
\boldmath
\qquad
\qquad
{\boldsymbol \theta}
\qquad
\qquad
\qquad
\qquad
\qquad
\qquad
\qquad
&
&
{\bf Families}
\nonumber \\
\label{cevode1}
\frac{1-\frac{\lambda }{q-r}}{S}
- \frac{\frac{q-r}{\alpha^{2}}}{S^{1+2 \beta}}
\quad \qquad \qquad \qquad
\qquad \qquad
&
&
\! \!\qquad {\rm F1}
\\
\label{cevode2}
\frac{1-\frac{\lambda }{q-r}}{S}
- \frac{\frac{q-r}{\alpha^{2}}}{S^{1+2 \beta}}
\quad \qquad \qquad \qquad
\qquad \qquad
&
&
\! \!\qquad {\rm F2}
\\
\label{cevode3}
\frac{2 \beta + \frac{\lambda}{q-r}}{S}
+ \frac{\frac{q-r}{\alpha^{2}}}{S^{1+2 \beta}}
\quad \qquad \qquad \qquad
\qquad \qquad
&
&
\! \!\qquad {\rm F3}
\\
\label{cevode4}
\frac{2 \beta + \frac{\lambda}{q-r}}{S}
+ \frac{\frac{q-r}{\alpha^{2}}}{S^{1+2 \beta}}
\quad \qquad \qquad \qquad
\qquad \qquad
&
&
\! \!\qquad {\rm F4}
\end{eqnarray}
The functional form of $\theta$ is the same
both
in
Families F1 and F2 (Equations (\ref{cevode1})  and (\ref{cevode2})) and in Families F3 and F4 (Equations (\ref{cevode3})  and (\ref{cevode4})).
However the functions $\theta$ are different
both in Families F1 and F2
and in Families F3 and F4
since $\lambda$  is different in all Families
F1, F2, F3, and F4 (Equations (\ref{ygflygflygflty}), (\ref{gravity7}), (\ref{gravity8}), and (\ref{gravity9})).



\noindent \bf {Fourth step - Output:} \normalfont

\noindent For n=1 Equations (\ref{karokaro}) imply \be
\label{asaasa} {\rm P}_{1}=-{\rm P}, \ee \be \label{asaasa1} {\rm
P}_{0}={\rm P}^{'}+\theta{\rm P}, \ee and \be
\label{gfkkukku} {\rm P}_{-1}=0={\rm P}^{''}+2\theta{\rm P}^{'}+ (
\theta^{2}+\theta^{'}-\nu  ) {\rm P}. \ee Combining Equations
(\ref{doladola}), (\ref{asaasa}) and (\ref{asaasa1}) yields \be
\label{asaasa3} \omega=\frac{{\rm P}^{'}}{{\rm P}} + \theta. \ee
For each of the retained families, i.e.
for each of the families F1, F2, F3, and F4,
we search for a polynomial P of
degree ${\rm d}$ (as defined in step \textbf{3a}) such that
Equation (\ref{gfkkukku}) is satisfied.
If such a
polynomial ${\rm P}$ is found then $\omega$ is given by Equation
(\ref{asaasa3}) and the function \be \label{dramdram}
\eta=\mathcal C=e^{\int \omega {\rm d}r}= e^{\int  ( \frac {{\rm
P}^{'}}{{\rm P}}+\theta  ){\rm d}r}= {\rm P} e^{\int \theta{\rm
d}r} \ee is a Liouvillian solution of Equation (\ref{cevodenew}).
Then the change of the dependent variable
(\ref{chvar})
$$
C=e^{-\frac {1}{2} \int \frac{2 (r-q)}{\alpha^{2} S^{2 \beta + 1} } \rm{d} \it{S} }
\mathcal C
$$
gives a Liouvillian solution to the CEV ODE
(\ref{redf}).


The application of the Fourth step of Kovacic's algorithm, when
$2 \beta=2,3,...$, is given in detail in Section \ref{classes}.
\indent


\noindent
Now the
algorithm can be considered complete when n=1. \newline

\vspace{-0.457cm}

\section{Classes of elementary function solutions
for half$-$integer values of $\beta$}
\label{classes}


\indent

For each of the retained families, i.e.
for each of the families F1, F2, F3, and F4,
we search for a polynomial solution P of
degree ${\rm d}$, as defined in step \textbf{3a}, such that
Equation (\ref{gfkkukku})
$$
{\rm P}^{''}+2\theta{\rm P}^{'}+ (
\theta^{2}+\theta^{'}-\nu  ) {\rm P}=0
$$
is satisfied.

The function $\nu$ is given by Equation (\ref{nnn})
$$
\nu=\frac{(q-r)^{2} + \alpha^{2}( (2 \beta-1) (q-r) +  2 \lambda
  )S^{2\beta}
}{\alpha^{4} S^{2+4\beta}}.
$$

Given that $\lambda$ in the
four families F1, F2, F3, and F4,
is given respectively by Equations
(\ref{ygflygflygflty}),
(\ref{gravity7}),
(\ref{gravity8}), and
(\ref{gravity9}),
the function $\nu$ in the four families
reads
\begin{eqnarray}
\boldmath
\qquad
\qquad
{\boldsymbol \nu}
\qquad
\qquad
\qquad
\qquad
\qquad
\qquad
\qquad
&
&
{\bf Families}
\nonumber \\
\label{nu1}
\frac{\left( \frac{q-r}{\alpha^{2}}   \right)^{2}}{S^{2 + 4 \beta}} + \frac{ \frac{
( 2 (\beta +  \rm n) - 1) \it (q-r)
}{\alpha^2}}{S^{2 + 2 \beta}}
\quad \qquad \qquad
\quad
\quad
\qquad
\; \;
&
&
\! \!\qquad {\rm F1}
\\
\label{nu2}
\frac{\left( \frac{q-r}{\alpha^{2}}   \right)^{2}}{S^{2 + 4 \beta}} + \frac{ \frac{
( 2 (\beta +  \rm n) + 1) \it (q-r)
}{\alpha^2}}{S^{2 + 2 \beta}}
\quad \qquad \qquad
\quad
\quad
\qquad
\; \;
&
&
\! \!\qquad {\rm F2}
\\
\label{nu3}
\frac{\left( \frac{q-r}{\alpha^{2}}   \right)^{2}}{S^{2 + 4 \beta}} - \frac{ \frac{
( 2 (\beta +  \rm n) - 1) \it (q-r)
}{\alpha^2}}{S^{2 + 2 \beta}}
\quad \qquad \qquad
\quad
\quad
\quad
\quad
\; \;
&
&
\! \!\qquad {\rm F3}
\\
\label{nu4}
\frac{\left( \frac{q-r}{\alpha^{2}}   \right)^{2}}{S^{2 + 4 \beta}} - \frac{ \frac{
( 2 (\beta +  \rm n) + 1) \it (q-r)
}{\alpha^2}}{S^{2 + 2 \beta}}
\quad \qquad \qquad
\quad
\quad
\quad
\quad
\; \;
&
&
\! \!\qquad {\rm F4}
\end{eqnarray}
and the function $\theta$ in the four families
reads
\begin{eqnarray}
\boldmath
\qquad
\qquad
{\boldsymbol \theta}
\qquad
\qquad
\qquad
\qquad
\qquad
\qquad
\qquad
&
&
{\bf Families}
\nonumber \\
\label{cevode1a}
\frac{1- \rm n
}{S}
- \frac{\frac{q-r}{\alpha^{2}}}{S^{1+2 \beta}}
\quad \qquad \qquad \qquad
\qquad \qquad
&
&
\! \!\qquad {\rm F1}
\\
\label{cevode2a}
- \frac{\rm n
}{S}
- \frac{\frac{q-r}{\alpha^{2}}}{S^{1+2 \beta}}
\quad \qquad \qquad \qquad
\qquad \qquad
&
&
\! \!\qquad {\rm F2}
\\
\label{cevode3a}
 \frac{1 - \rm n
}{S}
+ \frac{\frac{q-r}{\alpha^{2}}}{S^{1+2 \beta}}
\quad \qquad \qquad \qquad
\qquad \qquad
&
&
\! \!\qquad {\rm F3}
\\
\label{cevode4a}
- \frac{ \rm n
}{S}
+ \frac{\frac{q-r}{\alpha^{2}}}{S^{1+2 \beta}}
\quad \qquad \qquad \qquad
\qquad \qquad
&
&
\! \!\qquad {\rm F4}
\end{eqnarray}
where n is a non$-$negative integer.

With  the function $\nu$ given in the four families F1, F2, F3, and F4,  by
Equations (\ref{nu1}), (\ref{nu2}),
(\ref{nu3}), and (\ref{nu4}) respectively,
and the function $\theta$ given in the four families F1, F2, F3, and F4,  by
Equations (\ref{cevode1a}), (\ref{cevode2a}),
(\ref{cevode3a}), and (\ref{cevode4a}) respectively, Equation   (\ref{gfkkukku})
in the four families F1, F2, F3, and F4
reads

\begin{eqnarray}
\label{eqone}
{\rm P}^{''}+2  \left (      \frac{1- \rm n
}{S}
- \frac{\frac{q-r}{\alpha^{2}}}{S^{1+2 \beta}} \right )          {\rm P}^{'}+
\frac{\rm n(n-1)}{\it S^{\rm 2}}
{\rm P}=0&& \rm  F1
\\
\label{eqtwo}
{\rm P}^{''}+2  \left (      - \frac{\rm n
}{S}
- \frac{\frac{q-r}{\alpha^{2}}}{S^{1+2 \beta}} \right )          {\rm P}^{'}+
\frac{\rm n(n+1)}{\it S^{\rm 2}}
{\rm P}=0&& \rm  F2 \\
\label{eqthree}
{\rm P}^{''}+2  \left (       \frac{1 - \rm n
}{S}
+ \frac{\frac{q-r}{\alpha^{2}}}{S^{1+2 \beta}} \right )          {\rm P}^{'}+
\frac{\rm n(n-1)}{\it S^{\rm 2}}
{\rm P}=0 && \rm F3 \\
\label{eqfour}
{\rm P}^{''}+2  \left (       - \frac{ \rm n
}{S}
+ \frac{\frac{q-r}{\alpha^{2}}}{S^{1+2 \beta}} \right )          {\rm P}^{'}+
\frac{\rm n(n+1)}{\it S^{\rm 2}}
{\rm P}=0 && \rm F4
\end{eqnarray}

We search for polynomial solutions P of degree n,
where n is a non$-$negative integer, to the Equations (\ref{eqone}), (\ref{eqtwo}), (\ref{eqthree}), and (\ref{eqfour}). It is appropriate at this point to recall the following definition:
The function
\be
F(e,q;u)=\sum_{k=0}^{\infty} \frac{(e)_{k}}{(q)_{k}} \frac{u^{k}}{k!},
\ee
where the symbol
$(w)_{k}$, is the  Pochammer's symbol, and is defined by
\begin{equation}
(w)_{k}=w (w+1) ... (w+k-1),
\end{equation}
is called confluent hypergeometric function of the first kind or Kummer's function of the first kind.
When $e=-\rm m$, $\rm m$  being
a non$-$negative integer,
$F(e,q;u)$ is truncated and it reduces to a polynomial ${\rm P}_{\rm m}(u)=F(-\rm {m}$$,q;u)$  of degree $\rm m$
\be
\label{polsol}
F(-\rm {m}, \it q;u)=\rm 1- \frac{\rm m}{\it q} {\it u} + \frac{\rm m (\rm m -1)} {\it q (\it q+\rm 1)} \frac{{\it u}^{\rm 2}}{\rm 2!}+...+\frac{(-1)^{\rm m} \rm m !}{\it q(\it q+\rm 1)...(\it q+\rm m -1)} \frac{{\it u}^{\rm m}}{\rm m!}.
\ee

We easily find that the two$-$dimensional solution spaces of Equations (\ref{eqone}), (\ref{eqtwo}),
(\ref{eqthree}), and (\ref{eqfour}) are spanned
respectively by the following four pairs of functions

\vspace{0.2cm}
\noindent
\hspace{6cm}  $\pmb {1^{\rm st} \ \rm pair
}$
\normalfont
\begin{eqnarray}
\label{fampol1}
f_{1}(S)&=& S^{\rm n} F \left (-\frac{\rm n}{2 \beta},1-\frac{1}{2 \beta}; \frac{r-q}{a^{2}\beta S^{2 \beta}} \right )  \\
\label{fampol2}
f_{2}(S)&=& S^{\rm n - 1} F \left (-\frac{\rm n - 1}{2 \beta},1+\frac{1}{2 \beta}; \frac{r-q}{a^{2}\beta S^{2 \beta}} \right )
\end{eqnarray}
We note that $f_{1}(S)$ truncates when $\rm n$
is a multiple of $2 \beta$ and
becomes a polynomial of degree n, and
$f_{2}(S)$ truncates when $\rm n - 1$
is a multiple of $2 \beta$, $\beta=1, \frac{3}{2}, 2, \frac{5}{2},... ,\ $ and it becomes a polynomial of
degree $\rm n - 1$.
$f_{1}(S)$ and $f_{2}(S)$ span the solution space of
Equation (\ref{eqone}).

\vspace{0.2cm}
\noindent
\hspace{6cm}  $\pmb {2^{\rm nd} \ \rm pair
}$
\normalfont
\begin{eqnarray}
\label{fampol3}
f_{3}(S)&=& S^{\rm n} F \left (-\frac{\rm n}{2 \beta},1+\frac{1}{2 \beta}; \frac{r-q}{a^{2}\beta S^{2 \beta}} \right )  \\
\label{fampol4}
f_{4}(S)&=& S^{\rm n + 1} F \left (-\frac{\rm n + 1}{2 \beta},1-\frac{1}{2 \beta}; \frac{r-q}{a^{2}\beta S^{2 \beta}} \right )
\end{eqnarray}
We note that $f_{3}(S)$ truncates when $\rm n$
is a multiple of $2 \beta$ and
becomes a polynomial of degree n, and $f_{4}(S)$ truncates when $\rm n + 1$
is a multiple of $2 \beta$, $\beta=1, \frac{3}{2}, 2, \frac{5}{2},... ,\ $ and it becomes a polynomial of
degree $\rm n + 1$.
$f_{3}(S)$ and $f_{4}(S)$ span the solution space of
Equation (\ref{eqtwo}).

\vspace{0.2cm}
\noindent
\hspace{6cm}  $\pmb {3^{\rm nd} \ \rm pair
}$
\normalfont
\begin{eqnarray}
\label{fampol5}
f_{5}(S)&=& S^{\rm n} F \left (-\frac{\rm n}{2 \beta},1-\frac{1}{2 \beta}; \frac{q-r}{a^{2}\beta S^{2 \beta}} \right )  \\
\label{fampol6}
f_{6}(S)&=& S^{\rm n - 1} F \left (-\frac{\rm n - 1}{2 \beta},1+\frac{1}{2 \beta}; \frac{q-r}{a^{2}\beta S^{2 \beta}} \right )
\end{eqnarray}
We note that $f_{5}(S)$ truncates when $\rm n$
is a multiple of $2 \beta$ and
becomes a polynomial of degree n, and $f_{6}(S)$ truncates when $\rm n - 1$
is a multiple of $2 \beta$, $\beta=1, \frac{3}{2}, 2, \frac{5}{2},... ,\ $ and it becomes a polynomial of
degree $\rm n - 1$.
$f_{5}(S)$ and $f_{6}(S)$ span the solution space of
Equation (\ref{eqthree}).

\vspace{0.2cm}
\noindent
\hspace{6cm}  $\pmb {4^{\rm th} \ \rm pair
}$
\normalfont
\begin{eqnarray}
\label{fampol7}
f_{7}(S)&=& S^{\rm n} F \left (-\frac{\rm n}{2 \beta},1+\frac{1}{2 \beta}; \frac{q-r}{a^{2}\beta S^{2 \beta}} \right )  \\
\label{fampol8}
f_{8}(S)&=& S^{\rm n + 1} F \left (-\frac{\rm n + 1}{2 \beta},1-\frac{1}{2 \beta}; \frac{q-r}{a^{2}\beta S^{2 \beta}} \right )
\end{eqnarray}
We note that $f_{7}(S)$ truncates when $\rm n$
is a multiple of $2 \beta$ and
becomes a polynomial of degree n, and $f_{8}(S)$ truncates when $\rm n + 1$
is a multiple of $2 \beta$, $\beta=1, \frac{3}{2}, 2, \frac{5}{2},... ,\ $ and it becomes a polynomial of
degree $\rm n + 1$.
$f_{7}(S)$ and $f_{8}(S)$ span the solution space of
Equation (\ref{eqfour}).

\vspace{0.2cm}

Therefore we do get polynomial solutions to Equations
(\ref{eqone}), (\ref{eqtwo}),
(\ref{eqthree}), and (\ref{eqfour}), two for each equation, provided that n, n$-$1, or n+1 is a
multiple of 2$\beta$, depending on the family under consideration. In fact in this case each $f_{i}(S)$,
$i=1,2,...,8,$ represents a family of polynomials. For
example $f_{1}(S)$ truncates to a polynomial of degree n,
for \it each \normalfont value of n in the set $ \{ 2 \beta, 4 \beta, 6 \beta, ...     \},$ $ \beta=1, \frac{3}{2}, 2, \frac{5}{2},... \ .$ Hereafter when we refer to $f_{i}(S)$, $i=1,2,...,8,$ we will mean the associated denumerably infinite 
family of polynomials.
Since we look for
polynomial solutions of degree n
to Equations
(\ref{eqone}), (\ref{eqtwo}),
(\ref{eqthree}), and (\ref{eqfour}),
we conclude that
there remain four families we have to consider, namely the families
$f_{1}(S), f_{3}(S), f_{5}(S),$ and  $f_{7}(S).$





By combining Equations (\ref{sepcev}), (\ref{odecev}),
(\ref{chvar}),(\ref{ygflygflygflty}), (\ref{gravity7}), (\ref{gravity8}), (\ref{gravity9}), and
(\ref{dramdram}), we obtain that the four families of
polynomial solutions  $f_{1}(S), f_{3}(S), f_{5}(S),$ and  $f_{7}(S),$ to Equations (\ref{eqone}), (\ref{eqtwo}),
(\ref{eqthree}), and (\ref{eqfour}), give rise respectively to the following four classes of elementary
function solutions to the CEV PDE (\ref{CEVPAR})

\vspace{0.2cm}
\noindent
\hspace{6cm}  $\pmb {1^{\rm st} \ \rm class
}$
\normalfont
\begin{eqnarray}
\label{class1}
\mathcal S_{1,\rm n}(S)&=& S F \left (-\frac{\rm n}{2 \beta},1-\frac{1}{2 \beta}; \frac{r-q}{a^{2}\beta S^{2 \beta}} \right ) e^{- \rm n \it (q - r) t},
\end{eqnarray}
where $\rm n$
is a multiple of $2 \beta$, $\beta=1, \frac{3}{2}, 2, \frac{5}{2},... \ .$ The first class arises from the
family of polynomials $f_{1}(S)$ given by Equation (\ref{fampol1}). We
easily check that for n=0 we also obtain an elementary
function solution to the CEV PDE (\ref{CEVPAR}).


\vspace{0.2cm}
\noindent
\hspace{6cm}  $\pmb {2^{\rm nd} \ \rm class
}$
\normalfont
\begin{eqnarray}
\label{class2}
\mathcal S_{2,\rm n}(S)&=&
F \left (-\frac{\rm n}{2 \beta},1+\frac{1}{2 \beta}; \frac{r-q}{a^{2}\beta S^{2 \beta}} \right ) e^{-(\rm n + 1) \it (q-r)t},
\end{eqnarray}
where $\rm n$
is a multiple of $2 \beta$, $\beta=1, \frac{3}{2}, 2, \frac{5}{2},... \ .$ The second class arises from the
family of polynomials $f_{3}(S)$ given by Equation (\ref{fampol3}). We
easily check that for n=0 we also obtain an elementary
function solution to the CEV PDE (\ref{CEVPAR}).


\vspace{0.2cm}
\noindent
\hspace{6cm}  $\pmb {3^{\rm nd} \ \rm class
}$
\normalfont
\begin{eqnarray}
\label{class3}
\mathcal S_{3,\rm n}(S)&=&
e^{\frac{r-q}{a^{2}\beta S^{2 \beta}}}
S  F \left (-\frac{\rm n}{2 \beta},1-\frac{1}{2 \beta}; \frac{q-r}{a^{2}\beta S^{2 \beta}} \right ) e^{(2 \beta + \rm n -1) \it (q-r)t},
\end{eqnarray}
where $\rm n$
is a multiple of $2 \beta$, $\beta=1, \frac{3}{2}, 2, \frac{5}{2},... \ .$ The third class arises from the
family of polynomials $f_{5}(S)$ given by Equation (\ref{fampol5}). We
easily check that for n=0 we also obtain an elementary
function solution to the CEV PDE (\ref{CEVPAR}).


\vspace{0.2cm}
\noindent
\hspace{6cm}  $\pmb {4^{\rm th} \ \rm class
}$
\normalfont
\begin{eqnarray}
\label{class4}
\mathcal S_{4,\rm n}(S)&=& e^{\frac{r-q}{a^{2}\beta S^{2 \beta}}} F \left (-\frac{\rm n}{2 \beta},1+\frac{1}{2 \beta}; \frac{q-r}{a^{2}\beta S^{2 \beta}} \right )
e^{(2 \beta + \rm n) \it (q-r)t},
\end{eqnarray}
where $\rm n$
is a multiple of $2 \beta$, $\beta=1, \frac{3}{2}, 2, \frac{5}{2},... \ .$ The fourth class arises from the
family of polynomials $f_{7}(S)$ given by Equation (\ref{fampol7}). We
easily check that for n=0 we also obtain an elementary
function solution to the CEV PDE (\ref{CEVPAR}).
This completes the consideration of the problem when $2\beta=1,2,3, ... \ .$
\vspace{0.1cm}

There remain four cases to consider
$2 \beta=1,$ $2 \beta=0,$ $2 \beta=-1,$
and $2 \beta=-2,-3,...$
(Equation(\ref{per})).
The implementation of Kovacic's algorithm
in each one of these cases is by no means
trivial and it poses its own problems and challenges
\cite{melas}. The results in all cases are summarized in the following Theorem.




\begin{thrm}
\label{solutions}
The CEV PDE (Equation (\ref{CEVPAR}))
$$
\frac{\partial V}{\partial t} + \frac{1}{2}
\alpha^{2} S^{2 \beta + 2} \frac{\partial^{2} V}{\partial S^{2}} + (r-q) S \frac{\partial V}{\partial S} - (r-q) V =0,
$$
\begin{enumerate}
\item{
When $\beta=...,-\frac{5}{2},-2,-\frac{3}{2},-1,  1, \frac{3}{2}, 2, \frac{5}{2},... $ admits the following four classes of
elementary function solutions
\begin{eqnarray}
\mathcal S_{1,\rm n}(S) & = &S \ F \left (-\frac{\rm n}{2 \beta},1-\frac{1}{2 \beta}; \frac{r-q}{a^{2}\beta S^{2 \beta}} \right ) e^{- \rm n \it (q - r) t}, \nonumber \\
\mathcal S_{2,\rm n}(S)&=&
F \left (-\frac{\rm n}{2 \beta},1+\frac{1}{2 \beta}; \frac{r-q}{a^{2}\beta S^{2 \beta}} \right ) e^{-(\rm n + 1) \it (q-r)t}, \nonumber \\
\mathcal S_{3,\rm n}(S)&=&
e^{\frac{r-q}{a^{2}\beta S^{2 \beta}}} \
S  \ F \left (-\frac{\rm n}{2 \beta},1-\frac{1}{2 \beta}; \frac{q-r}{a^{2}\beta S^{2 \beta}} \right ) e^{(2 \beta + \rm n -1) \it (q-r)t}, \nonumber \\
\mathcal S_{4,\rm n}(S)&=& e^{\frac{r-q}{a^{2}\beta S^{2 \beta}}} F \left (-\frac{\rm n}{2 \beta},1+\frac{1}{2 \beta}; \frac{q-r}{a^{2}\beta S^{2 \beta}} \right )
e^{(2 \beta + \rm n) \it (q-r)t}, \nonumber
\end{eqnarray}
where \rm n= \rm 0, \normalfont \it or \rm n \it is any multiple of \rm 2$\beta$,
\item{\it When $\beta=\frac{1}{2}$
admits the classes $\mathcal S_{2,\rm n}(S)$
and $\mathcal S_{4,\rm n}(S)$,
where \rm n= \rm 0, \normalfont \it or \rm n \it is any positive integer,
}
\item{\it When $\beta=-\frac{1}{2}$
admits the classes $\mathcal S_{1,\rm n}(S)$
and $\mathcal S_{3,\rm n}(S)$,
where \rm n= \rm 0, \normalfont \it or \rm n \it is any negative integer,
}

\item{\it When $\beta=0$
admits the elementary function
solution
$$
e^{- \lambda t} \left ( g S^{\frac{\xi_{1}+\zeta_{1}}{2 \sigma^{2}}} + h S^{\frac{\xi_{\rm 1}-\zeta_{\rm 1}}{\rm 2 \sigma^{\rm 2}}} \right ),
$$
\rm where,
$$
 \xi_{1}= \sigma^{\rm{2}} - \rm{2} ( \it{r-q}), \ \ \zeta_{\rm 1}=\sqrt{(\sigma^{\rm{2}} - \rm{2} (\it{r-q}))^{\rm{2}} + \rm{8} \sigma^{2}(\it{r-q}+\lambda)},
$$
and
$g, \ h,$ and $\lambda$ are arbitrary real numbers
(Equations (\ref{coefficients}), (\ref{liou2})).
}}
\end{enumerate}
\end{thrm}

Since the CEV PDE is linear an immediate
Corollary of Theorem \ref{solutions} is
the following

\begin{cor}
The CEV PDE (Equation (\ref{CEVPAR}))
$$
\frac{\partial V}{\partial t} + \frac{1}{2}
\alpha^{2} S^{2 \beta + 2} \frac{\partial^{2} V}{\partial S^{2}} + (r-q) S \frac{\partial V}{\partial S} - (r-q) V =0,
$$
\begin{enumerate}
\item{
When $\beta=...,-\frac{5}{2},-2,-\frac{3}{2},-1,  1, \frac{3}{2}, 2, \frac{5}{2},... $ admits the following
elementary function solution
\be
\mathcal F_{1}(S) =
\sum_{k=1}^{q} m_{k} \mathcal S_{1,\mu_{\it k}}(S) +
\sum_{k=1}^{r} z_{k} \mathcal S_{2,\zeta_{\it k}}(S)+
\sum_{k=1}^{t} w_{k} \mathcal S_{3,\omega_{\it k}}(S)+
\sum_{k=1}^{j} d_{k} \mathcal S_{4,\delta_{\it k}}(S),
\ee
where $q,r,t,j$ are any positive integers, equal to one or greater
than one, $m_{k},z_{k},w_{k}$ and
$d_{k}$ are arbitrary real numbers,
and $\mu_{\it k},\zeta_{\it k},\omega_{\it k}$ and $\delta_{\it k}$
are any multiples of $2 \beta$ or 0,}

\item{\it When $\beta=\frac{1}{2}$
admits the  following
elementary function solution
\be
\mathcal F_{2}(S) =
\sum_{k=1}^{g} e_{k} \mathcal S_{2,\varepsilon_{\it k}}(S)+
\sum_{k=1}^{h} y_{k} \mathcal S_{4,\upsilon_{\it k}}(S),
\ee
where $g$ and $h$ are any positive integers, equal to one or greater
than one, $e_{k}$ and
$y_{k}$ are arbitrary real numbers,
and, $\varepsilon_{\it k}$ and $\upsilon_{\it k}$
are any positive integers or 0,}

\item{\it When $\beta=-\frac{1}{2}$
admits the  following
elementary function solution
\be
\mathcal F_{3}(S) =
\sum_{k=1}^{c} p_{k} \mathcal S_{1,\pi_{\it k}}(S)+
\sum_{k=1}^{l} r_{k} \mathcal S_{3,\varrho_{\it k}}(S),
\ee
where $c$ and $l$ are any positive integers, equal to one or greater
than one, $p_{k}$ and
$r_{k}$ are arbitrary real numbers,
and, $\pi_{\it k}$ and $\varrho_{\it k}$
are any negative integers or 0.}

\end{enumerate}
\end{cor}

\vspace{0.2cm}

\indent
The following three remarks are in order
regarding the derived classes of elementary
function solutions to the CEV model.
\begin{enumerate}
\item{The class $\mathcal S_{1,\rm n}(S)$, given by Equation (\ref{class1}), is contained implicitly in the solution $C_{1}(S)$ given by Equation (\ref{c1}),
the class
$\mathcal S_{2,\rm  n}(S)$,
given
by Equation (\ref{class2}),
is
contained implicitly in the solution $C_{2}(S)$ given by Equation (\ref{c2}),
  whereas the classes $\mathcal S_{3,\rm n}(S)$ and $\mathcal S_{4,\rm n}(S)$,
given respectively by Equations (\ref{class3}) and (\ref{class4}),
are not
contained in the solutions $C_{1}(S)$ and $C_{2}(S).$}

\item{When $\beta=0$
the general solution of the CEV ODE is
Liouvillian (Equations (\ref{coefficients}), (\ref{liou2})).
Thus it is no surprise that in this case
Kovacic's algorithm gives the two Liouvillian solutions
$S^{\frac{\xi_{1}+\zeta_{1}}{2 \sigma^{2}}}$ and $S^{\frac{\xi_{\rm 1}-\zeta_{\rm 1}}{\rm 2 \sigma^{\rm 2}}}$
which span the
solution space of the CEV ODE.
}

\item{It is well known (Equation (\ref{kritskrits})) that for every Liouvillian solution $ f
     (S)$ 
    to the CEV ODE (Equation (\ref{ODECEV}))
$$
\frac{1}{2}
\alpha^{2} S^{2 \beta + 2} \frac{\rm{d}^{2} \it{ C}}{\rm{d} \it{S}^{\rm{2}}} + (r-q) S \frac{\rm{d} \it{ C}}{ \rm{d} \it{S}} - (r-q+\lambda) C =0
$$
there is a second Liouvillian solution $
L(S)=f
(S) \int \frac { e^{ - \int \frac {\rm 2 \it (r-q)}{
\alpha^{\rm 2} S^{\rm 2 \beta + 1}
} \rm{d} \it{S}
}}{ f
(S) ^{2}} \rm{d} \it{S}
$
to the CEV ODE which also gives rise
(Equation (\ref{sepcev})) to a
solution $\mathcal L(S)=f
(S) \int \frac { e^{ - \int \frac {\rm 2 \it (r-q)}{
\alpha^{\rm 2} S^{\rm 2 \beta + 1}
} \rm{d} \it{S}
}}{ f
(S) ^{2}} \rm{d} \it{S}  e^{- \lambda
t}$ to the CEV PDE (Equation(\ref{CEVPAR}))
$$
\frac{\partial V}{\partial t} + \frac{1}{2}
\alpha^{2 } S^{2 \beta + 2} \frac{\partial^{2} V}{\partial S^{2}} + (r-q) S \frac{\partial V}{\partial S} - (r-q) V =0.
$$

In all cases the integral $   \int \frac { e^{ - \int \frac {\rm 2 \it (r-q)}{
\alpha^{\rm 2} S^{\rm 2 \beta + 1}
} \rm{d} \it{S}
}}{ f
(S) ^{2}} \rm{d} \it{S} $
 is nonelementary and it can
only be
 evaluated by using Taylor's series, or
numerically, by using
quadrature rules such as the Simson's rule or
Gaussian rule. Consequently some kind of
approximation scheme is needed in order to extract information from the solutions
$\mathcal L(S)$ for the pricing of the financial instruments they describe.
For this reason we do not include the
solutions $\mathcal L(S)$ to the classes
of elementary function solutions to the
CEV model we derive in this paper.


}



\end{enumerate}

\section{Conclusion and Future Development}
\label{future}

\indent

Analytical tractability of any financial model is an important feature. Existence of
a closed$-$form solution definitely helps in pricing financial instruments and calibrating
the model to market data. It also helps to verify the model assumptions, check its
asymptotic behavior and explain causality. In fact, in mathematical finance many models were
proposed, first based on their tractability, and only then by making another argument.

If the true asset price
process was geometric Brownian motion with constant
volatility, then the BSM PDE could
be used to find out this volatility by equating the
model price of a standard option to its market price (implied volatility).

Empirically, we find that the
implied volatilities computed from market prices of
options with different strike prices are not constant
but vary with strike price. This variation is observed
across a wide range of markets and underlying assets
and is known as the implied volatility smile or frown
depending on its shape. The lognormal assumption
with constant volatility of the BSM model does not capture this effect. The CEV model
is capable of reproducing
the volatility smile observed in the empirical data.

Our
classes of elementary function solutions
to the CEV model
allow fast and accurate calculation of prices of various financial instruments
under the CEV process. Moreover
they will facilitate further the use of the CEV model as a benchmark for
the pricing of
various types of financial instruments.

In future research we will study the Lie point
symmetries of the CEV PDE and we will use
the classes of elementary function solutions
to the CEV PDE we derived in this paper
in order to obtain more
solutions to the CEV model in terms of elementary
functions. The study of the financial instruments
which they describe  will be useful
 for trading by using CEV diffusion in all cases.



\section*{APPENDIX}
\label{appendix}

\it{Notations.} \normalfont Let ${\rm
L}_{\rm{max}}=\{1,2,4,6,12\}$ and let $h$ be the function defined
on ${\rm L}_{\rm{max}}$ by \be
h(1)=1\,\,,\,\,h(2)=4,\,\,\,\,h(4)=h(6)=h(12)=12. \ee

\noindent \bf{Input:} \normalfont A rational function
$$ \nu(x)=\frac{s(x)}{t(x)} \qquad ({\rm Equation} \\ (\ref{katrouba1})). $$
\noindent The polynomials $ s,t \in C[x]$ are supposed to be
relatively prime. \newline The differential equation under consideration is
$$ z''- \nu z= 0 \qquad ({\rm Equation} \\ (\ref{katrouba})).  $$

\noindent \bf {First step:} \normalfont { \it The set L of
possible degrees of $ \omega .$ } \normalfont

\noindent We are interested in Equation (\ref{katrouba}) where
$\nu(x)$ is given by (\ref{katrouba1}).

\noindent \textbf{1a}. \hspace{0.5 cm} If $t(x)=1$, set $m=0$.
Otherwise, factorize $t(x)$ \be
t=t_{1}t_{2}^{2}t_{3}^{3}...t_{m}^{m} \ee where the $t_{{\rm
i}}\,,\,{\rm i}=1,2,...,m,$ \,are relatively prime two by two and
each $t_{{\rm i}}$ either is equal to one or has simple zeros. Let
\be \Gamma'=\{ c\in C\,,\,t(c)=0 \} \,\,\,\, {\rm and} \,\,\,\,
\Gamma=\Gamma'\cup \{\infty\}, \ee where $\cup$ denotes
set$-$theoretic union. Associate orders with the elements of
$\Gamma$: \be {\rm o}(c)={\rm i} \ee for all $c\in \Gamma'$ ,where
i is such that $t_{{\rm i}}(c)=0$, and
$$
{\rm o}(\infty)={\rm max}(0, 4+{\rm d}^{{\rm o}}s- {\rm d}^{{\rm
o}}t).
$$
Let \be m^{+}={\rm max}(m,{\rm o}(\infty)). \ee For \,\,
$0\leq{\rm i}\leq m^{+}$ \,\, let \be \Gamma_{{\rm i}}=\{ c \in
\Gamma|\,{\rm o}(c)={\rm i} \}. \ee
\noindent \textbf{1b}. \hspace{0.5 cm} If \, $m^{+}\geq 2$ \,
define \,$\gamma_{2}$\, and \,$\gamma$ \,by \be \gamma_{2}=\left
|\Gamma_{2} \right|\,\,{\rm and}\,\, \gamma=\gamma_{2}+ \left |
\cup \Gamma_{\rm k} \right |, \,\, {\rm k}\,\,{\rm is}\,\,{\rm
odd},\,\,{\rm and}\,\, 3\leq{\rm k}\leq{\rm m}^{+} \ee
respectively, where, if S is a set then $|{\rm S}|$ denotes the
number of elements of S.


 $\!\!\!\!\!\!\!\!\!$ \textbf{1c}. \hspace{0.5 cm} Construct L, a
subset of
${\rm L}_{\rm{max}},$
as follows
\be
1\in {\rm L}\Longleftrightarrow \gamma=\gamma_{2}, \ee \be 2\in
{\rm L}\Longleftrightarrow \gamma\geq2\,\,,\,\,{\rm and}, \ee \be
4,6\,\,{\rm and}\,\,12 \in {\rm L}\Longleftrightarrow m^{+}\leq2.
\ee
\textbf{1d}. \hspace{0.5 cm} If \,${\rm L}=\emptyset$ \,go to the
stage of the algorithm END. Otherwise, let n  be equal to the
smallest element of L.

\noindent  \bf {Second step:}  \normalfont { \it
The  sets ${\rm E}_{c}$ associated to the singular points.}
\normalfont

\noindent  Construction of the sets ${\rm E}_{c}$, \,$c \in
\Gamma$.

$\!\!\!\!\!\!\!\!\!\!\!$
\textbf{2a}. \hspace{0.5 cm}
If $\infty \in \Gamma_{0}$ then E$_{\infty}=h(\rm n)\{0,1,...,\rm n\}.$


$\!\!\!\!\!\!$ \textbf{2b}. \hspace{0.5 cm} When
n=1, for each $c \in \Gamma_{2{\rm q}}$ with ${\rm q} \geq 2$,
calculate one of the two ``square roots'' $ \left [ \sqrt \nu
\,\right ] _c$ of $\nu$ defined as follows:

\noindent If $c \in C$, \be \label{MilkyWay} \left [ \sqrt \nu
\,\right ] _{c}=
\frac{{\rm a}_{c}}{(x-c)^{\rm q}}+
\sum_{{\rm i}={\rm q}-1}^{2} \frac{\mu_{{\rm i},c}}{(x-c)^{{\rm i}}},
\ee
and
\be \label{MilkyWay1} \nu-\left [ \sqrt \nu \,\right ]
_{c}^{2} = \frac{\rm b_{c}}{(x-c)^{{\rm q}+1}}
+ {\rm O} \left ( \frac{1}{(x-c)^{{\rm q}}}      \right ),\ee
where
the O symbol has its usual meaning, $ f(x)={\rm O}(g(x))$ if
$|f(x)|\leq{\rm M}g(x)$ for some positive constant M; $g(x)$ is assumed to be positive.

\noindent Let
\be \label{sauvage} {\rm E}_{c}=\left \{ \frac
{1}{2} \left ( {\rm q}+\epsilon \frac{{\rm b}_{c}}{{\rm
a}_{c}} \right ) \left | \right. \ \epsilon= \underline{+} 1 \right \}. \ee


\noindent Define a function ``Sign'' S with domain ${\rm
E}_{c}$ as follows \be \label{blooo} {\rm S} \left (
\frac{1}{2} \left ( {\rm q}+ \epsilon \frac{{\rm b}_{c}}{{\rm
a}_{c}} \right ) \right )=
 \left \{ \begin{array}{lr}
\epsilon &  {\rm if} \ \  {\rm b}_{c} \neq 0 \\
1 & {\rm otherwise}
\end{array}
\right . . \ee


\noindent
In Equation (\ref{MilkyWay}) $\left [ \sqrt \nu \,
\right ] _{c}$ is the sum of terms involving $(x-c)^{-{\rm i}}$
for $2 \leq {\rm i} \leq {\rm q}$ in the Laurent series for
$\sqrt \nu$ at $c$. In practice, one would not form the
Laurent series for $\sqrt \nu$, but rather would determine $ \left
[ \sqrt \nu \right ]_{c}$ by using undetermined coefficients,
i.e. by equating $\left ( \left [ \sqrt \nu \right ]_{c}
\right )^{2}= \left ( {\rm a}_{c}  (x-c)^{-{\rm q }} +       {\mu}_{{\rm q}-1,c} (x-c)^{-({\rm q }-1)} +... \right . $
$ \left . + \mu_{2, c} (x-c)^{-2}
\right )^{2}$ with the corresponding part of the Laurent series
expansion of $\nu$ at $c$. There are two possibilities for $
\left [ \sqrt \nu \right ]_{c}$, one being the negative of
the other, and any of these may be chosen. In Equation
(\ref{MilkyWay1}), $\nu$ denotes its Laurent series expansion at
$c$. This equation defines ${\rm b}_{c}$.

\noindent \bf {Third step:} \normalfont {\it Possible degrees for
P and possible values for $\theta$.} \normalfont

\noindent  $\!$ \textbf{3a}. \hspace{0.5 cm} For each family
$\underline{{\rm e}}= \left ({\rm e}_{c} \right ) _{c \in \Gamma}$
of elements ${\rm e}_{c} \in {\rm E}_{c}$ calculate \be {\rm d}
\left ( \underline{{\rm e}} \right )= {\rm n}- \frac{{\rm n}}{h
({\rm n} )} \sum_{c \in \Gamma} {\rm e}_{c} \ee \noindent
$\!$ \textbf{3b}. \hspace{0.5 cm}
Retain the families
$\underline{{\rm e}}$ which are such that
$$
\,\,\, {\rm d}(\underline{{\rm e}}) \in N \,\, {\rm where}
\, \,N \, \, {\rm \, \, is \, \, the \, \, set \, \, of \, \,
non-negative \, \, integers \, \, and}
$$
If none of the families $\underline{{\rm e}}$ is retained, go to
the stage of the algorithm $\underline{{\rm CONTINUATION}}$.

\noindent
$\!$ \textbf{3c}. \hspace{0.6 cm} For each family $\underline{{\rm
e}}$ retained from step \textbf{3b}, form the rational function
\be \label{krookraa} \theta=\frac{{\rm n}}{h({\rm n})}\sum_{c\in
\Gamma'} \frac{{\rm e}_{c}}{x-c} \,+ \, \delta_{{\rm n}}^{1}
\!\!\!\!\!\!\!\!\! \sum_{\begin{array}{c}
c \in \cup \Gamma_{2{\rm q}} \\
{\rm q} \geq 2
\end{array}}
\!\!\!\!\!\!\!\!\!{\rm S} \left ({\rm e}_{c} \right ) \left [
\sqrt r \right ]_{c}, \ee where $\delta_{{\rm n}}^{1}$ is the
Kronecker symbol.

\noindent  {\bf Fourth Step:} \normalfont {\it Tentative
computation of P.} \normalfont


\noindent Search for a polynomial P of degree d (as defined in
step \textbf{3a}) such that \be \label{karokaro}
\begin{array}{l}
{\rm P}_{{\rm n}}=-{\rm P} \\
\cdot \, \cdot \, \cdot \\
{\rm P}_{{\rm i}-1}=-{\rm P}^{'}_{{\rm i}} -\theta{\rm P}_{{\rm
i}}-
({\rm n}-{\rm i})({\rm i}+1)\nu{\rm P}_{{\rm i}+1} \\
\cdot \, \cdot \, \cdot \\
{\rm P}_{-1}=0,
\end{array}
\ee where the prime in ${\rm P}^{'}(x)$ denotes differentiation
with respect to the independent variable $x$.

\noindent \bf {Output:} \normalfont

\noindent
OUTPUT1: If such a polynomial is found and $\omega$ is a solution
of the irreducible algebraic  equation
$$
\sum_{{\rm i}=0}^{{\rm n}} \frac{{\rm P}_{{\rm i}}(x)} {({\rm
k}-{\rm i})!} \omega^{{\rm i}}=0 \,\,\,\,\,\,\,\,\,\, \left ( {\rm
Equation} \,\,\,\,(\ref{doladola}) \right ),
$$
where the rational functions  ${\rm P}_{{\rm i}}(x)$ are defined
in (\ref{karokaro}), then the function $\eta=e^{\int \omega }$ is
a Liouvillian solution of the equation under consideration
$$z''=\nu z  \,\,\,\,\,\,\,\,\,\, ({\rm Equation}(\ref{katrouba})).$$
If no such polynomial is found for any family retained from step
\textbf{3b}, go to the stage of algorithm $\underline{{\rm
CONTINUATION}}.$

\noindent $\underline{{\rm CONTINUATION}}:$\,\, If ${\rm n}$ is
different from ${\rm the } \,\, {\rm largest}\,\, {\rm
element}\,\,{\rm of}\,\,{\rm L}$ then set n equal to the next (in
increasing order) element of L and go to \textbf{Step 2}.


\noindent  OUTPUT2: Equation $z''=\nu z$ has no Liouvillian
solutions.

\vspace{0.2cm}


{}



\begin{thebibliography}{10}

\bibitem{BS0} F. Black and M. Scholes, The valuation
of option contracts and a test of market efficiency’
Journal of Finance \textbf{27}  399$–$417 (1972)


\bibitem{BS} F. Black and M. Scholes, The Pricing of Options and Corporate Liabilities’
\emph{Journal of Political Economy} \textbf{81} 637$-$654 (1973)



\bibitem{M0} R. C. Merton, Theory of Rational Option Pricing,
\emph{ The Bell Journal of Economics and Management Science} \textbf{4} 141$-$183 (1973)



\bibitem{M} R. C. Merton, Continuous Time Finance Blackwell (1990)




\bibitem{G} G. Marsaglia, Evaluating the Normal Distribution \emph{Journal of Statistical Software
 } \textbf{11}  (2004)




\bibitem{Cox} J. Cox, ‘Notes on option pricing I: Constant elasticity of variance diffusions’ Working paper,
Stanford University (1975) (Reprinted in \emph{J. Portf. manage.} \textbf{22} 15$-$17 (1996))

\bibitem{Cox1} J. Cox   and S. Ross, ‘The valuation of options for alternative stochastic processes’ \emph{Journal of Financial Economics} \textbf{3} 145$-$166 (1976)

\bibitem{Ema} D. C. Emanuel  and J. D. MacBeth,  ‘Further results on the constant elasticity
of variance call option pricing model’ \emph{Journal of Financial and Quantitative Analysis}
\textbf{17} 533$-$554 (1982)


\bibitem{Sch}  M. Schroder,  Computing the constant elasticity of variance option pricing formula
\emph{Journal of Finance} \textbf{44} 211$-$219  (1989)


\bibitem{Kou}  S. G. Kou,  A Jump$-$Diffusion Model for Option Pricing \emph{Management Science} \textbf{48} 1086$-$1101 (2002)


\bibitem{Gat} J. Gatheral, E. P. Hsu, P. Laurence, C. Ouyang, and Tai$-$Ho Wang, Asymptotics of Implied Volatility in
Local Volatility Models \emph{Mathematical Finance} \textbf{12}  published online
(2010)

\bibitem{Y} K. C. Yuen,  H. Yang and K. L. Chu, Estimation in the Constant Elasticity of Variance Model \emph{British Actuarial Journal}  \textbf{7}
 275$-$292 (2001)


\bibitem{Far} R. W. Farebrother,  Algorithm AS 231: The Distribution of a Noncentral $\chi^{2}$ Variable
with Nonnegative Degrees of Freedom \emph{Applied Statistics}  \textbf{36} 402$-$405 (1987)


\bibitem{Pos} Harry O. Posten,  An Effective Algorithm for the Noncentral Chi$-$Squared Distribution
Function \emph{American Statistician}  \textbf{43} 261$-$263 (1989)


\bibitem{Di} Cherng G. Ding,  1992, Algorithm AS 275: Computing the Non$-$Central $\chi^{2}$ Distribution
Function \emph{Applied Statistics}  \textbf{41} 478$-$482 (1992)


\bibitem{Bab} L. Kn\"{u}˜sel  and B. Bablok,  Computation of the Noncentral Gamma Distribution, \emph{SIAM
Journal of Scientiflc Computing}  \textbf{17} 1224$-$1231 (1996)


\bibitem{Ben} D. Benton  and K. Krishnamoorthy, 2003, Computing Discrete Mixtures of Continuous
Distributions: Noncentral Chisquare, Noncentral t and the Distribution of the Square of
the Sample Multiple Correlation Coefficient, \emph{Computational Statistics and Data Analysis}
\textbf{43} 249$-$267 (2003)

\bibitem{Dyr}   S.    Dyrting,  Evaluating the Noncentral Chi$-$Square Distribution for the Cox$-$Ingersoll$-$Ross Process \emph{Computational Economics} \textbf{24} 35$-$50 (2004)




\bibitem{Neu} A. Neuberger, The log contract
\emph{Journal of Portfolio Management}
\textbf{20} (1994)


\bibitem{Neu1} A. Neuberger, The log contract
and other power contracts, chapter 7 in
\emph{The Handbook of exotic options: Instruments, Analysis and Appilcations}
McGraw$-$Hill (1996)


\bibitem{Bou} E. Boussard, Trading and Hedging Volatility, Presentation at
    Derivatives '97, Risk Conference, Brussels, February 1997









\bibitem{Kova}
J. Kovacic,
An algorithm for solving second order linear homogeneous
differential equations
\emph{J. Symb. Comp.} \textbf {2} 3$-$43 (1986)





\bibitem{Bac} L. Bachelier,  ‘Th\'{e}orie de la speculation’  \emph{3rd Annales scientifiques de l'\`{E}cole Normale Sup\v{e}\`{e}érieure} \textbf{17} 21$–$86 (1900)


























\bibitem{Den} P. Dennis and S. Mayhew,
‘Risk$-$neutral skewness: Evidence from stock options’  \emph{Journal of Financial and Quantitative Analysis} \textbf{37} 471$-$493   (2002)

\bibitem{Wu} G. Bekaert and G. Wu,
‘Asymetric volatility and risk in equity markets’
 \emph{Review
of Financial Studies} \textbf{13} 1$-$42   (2000)


\bibitem{Mac} J. D. MacBeth and L. J. Merville,
‘Tests of the Black$-$Scholes and Cox call option
valuation models  \emph{Journal of Finance}
\textbf{35} 285$-$301  (1980)


\bibitem{Dia} J. C. Dias and J. P. V. Nunes,
Pricing Real Options under the CEV Diffusion
\emph{Journal of Futures Markets}
\textbf{31}  230$-$250 (2011)









\bibitem{Dav} D. Davydov and V. Linetsky,   Pricing and Hedging Path$-$Dependent Options under the CEV Process \emph{Management Science} \textbf{47} 947$-$965 (2001)


\bibitem{Lin} V. Linetsky,  Lookback options and diffusion hitting times: a spectral expansion approach \emph{Finance and Stochastics}
\textbf{8} 373$–$398 (2004)


\bibitem{Davy}
D. Davydov  and V. Linetsky,   Pricing options on scalar diffusions: An eigenfunction expansion approach \emph{Operations Research} \textbf{51} 185$-$209 (2003)


\bibitem{Cos} M. Costabile,
On pricing lookback options under the CEV process
\emph{Decisions in Economics and Finance}
\textbf{29}  139$-$153 (2006)




\bibitem{Boy0}P. P. Boyle, Y.  Tian, and J. Imai,  Lookback options under the CEV process: a correction \emph{Journal of Financial and Quantitative Analysis Unpublished Appendixes, Notes, Comments, and Corrections http://depts.washington.edu/jfqa/} \normalfont (1999)








\bibitem{Boy}
P. P. Boyle  and Y. Tian,   Pricing lookback and barrier options under the CEV
process \emph{Journal of Financial and Quantitative Analysis} \textbf{34} 241$-$264 (1999)































































































\bibitem{Paul} P. Wilmott, J. Dewynne, and S. Howison, Option Pricing: Mathematical models and computation Oxford Financial Press, Oxford  (1993)



\bibitem{Shaw} W. T. Shaw,
Modelling Financial
Derivatives  with
Mathematica
Cambridge University Press   (1998)




\bibitem{Olver} P. J. Olver, Applications of
Lie Groups to Differential Equations,
Springer$-$Verlag (1993)

\bibitem{Beck}
S. Beckers, The Constant Elasticity of Variance Model and Its Implications for Option
Pricing
\emph{The Journal of Finance} \textbf {35}
661$-$673 (1980)




\bibitem{Handbook}
HANDBOOK OF MATHEMATICAL FUNCTIONS with Formulas, Graphs, and Mathematical Tables
Edited by M. Abramowitz and I. A. Stegun
Addison$-$Wesley, U.S.A., and Scientific Computing Service Ltd., Great Britain
(1962)



\bibitem{kovec} P. P. Boyle, W. Tian and
F. Guan, The Riccati Equation in Mathematical
Finance
\emph{J. Symbol. Comput.} \textbf {33}
343$-$355 (2002)

















\bibitem{LA}
I. Hern$\acute{\rm{a}}$ndez, C. Mateos, J. N.
Vald$\acute{\rm{e}}$s and A. F. Tenorio,
Lie Theory: Applications to problems in Mathematical Finance and Economics
\emph{Applied Mathematics and Computation}
\textbf{208} 446$-$452 (2009)











\bibitem{Galois} E. Galois,  M\`{e}moire sur les conditions de r\`{e}solubilit\`{e} des \`{e}quations par radicaux (Dated
6 January 1831)
\emph{Journal de Math\`{e}matiques pures et appliqu\`{e}es} \normalfont (Ed. J. Liouville) (1846)


\bibitem{Galois1} E. Galois,  Des \`{e}quations primitives qui sont soîubîes par radicaux
\emph{Journal de Math\`{e}matiques pures et appliqu\`{e}es} \normalfont (Ed. J. Liouville) (1846)


\bibitem{Galois2} E. Galois,  Lettre \`{a} Auguste Chevalier (Dated 29 May 1832) Published in \emph{Revue encyclop\'{e}dique}
(Sept. 1832)



















\bibitem{Lie1}
S. Lie, Die Grundlagen f\"{u}r die Theorie der unendlichen kontinuierlichen Transformationsgruppen.
1. Abhandlung \emph{Abh. Ges. Wiss. Leipzig} \normalfont (1891) reprinted in Lie, S.: Gesammelte Abhandlungen  \textbf{6} Leipzig/Oslo
300$-$330  (1927)



\bibitem{Lie2}
S. Lie, Untersuchungen \"{u}ber unendliche kontinuierliche Gruppen  \emph{Abh. Ges. Wiss. Leipzig} \normalfont  (1895) reprinted in Lie, S.: Gesammelte Abhandlungen  \textbf{6} Leipzig/Oslo
396$-$493  (1927)


\bibitem{Piccard}
E. Picard,  Trait\'{e} d'analyse \textbf{3} (Deuxieme Ed.) Gauthier$-$Villars
(1908)



\bibitem{Vessiot}
E. Vessiot,  Sur l'int\'{e}gration des \'{e}quations diff\'{e}rentielles lin\'{e}aires \emph{Annales Scientifiques de l'\'{E}cole Normale Sup\'{e}rieure} \normalfont 3 \textbf{9} 197$–$280 (1892)



\bibitem{Vessiot1}
E. Vessiot,   M\'{e}thodes d'int\'{e}gration \'{e}l\'{e}mentaires
\emph{Encyclop\'{e}die des sciences math\'{e}matiques pures et appliqu\'{e}es}  \textbf{3} Gauthier$-$Villars \& Teubner,  58$-$170  (1910)



\bibitem{Abel} N. H. Abel, M\'{e}moire sur une classe particuli\`{e}re d' \'{e}quations r\'{e}solubles alg\'{e}briquement
\emph{Journ. Crelle} \textbf{4} \normalfont (1829)


\bibitem{Cant} B. J. Cantwell, Introduction to Symmetry Analysis Cambridge University Press (2002)










\bibitem{Ou}
W. R. Oudshoorn and van der Put M., Lie symmetries and differential Galois groups of linear equations \emph{Math.
Comp.} \textbf{71}  349$-$361 (2002)







\bibitem{Ibr}
N. H. Ibragimov, A bridge between Lie symmetries
and Galois groups Published in \emph{Differential Equations: Geometry, Symmetries and Integrability}
\normalfont
Proc. 5th Abel Symposium, Troms\"{o}, Norway, Hune 17$-$22, 2008, Springer
159$-$172 (2009)



\bibitem{Int}
D. BL$\acute{A}$ZQUEZ$-$SANZ, J. J.
MORALES$-$RUIZ and J. A. WEIL,
Differential Galois Theory and Lie Symmetries
\emph{SIGMA} \textbf {11} 092 (2015)


\bibitem{kolchin1}
E. R. Kolchin, Algebraic matric groups and the Picard-Vessiot theory
of homogeneous linear ordinary differential equations \emph{Annals of Mathematics}
 \textbf {49} 1$-$42 (1948)




\bibitem{kolchin2}
E. R. Kolchin, Differential Algebra and Algebraic Groups Academic
Press
(1976)


\bibitem{Boul}
A. Boulanger, Contribution $\grave{\rm{a}}$ l'$\acute{\rm{e}}$tude des $\acute{\rm{e}}$quations lin$\acute{\rm{e}}$aires homog$\grave{\rm{e}}$enes int$\grave{\rm{e}}$grables alg$\grave{\rm{e}}$briquement \emph{Journal de l'$\acute{E}$cole Polytechnique, ´
Paris}
\textbf {4} 1$-$122 (1898)




\bibitem{Singer2}
M. F. Singer, Liouvillian solutions of nth order homogeneous linear
differential equations
\emph{American Journal of Mathematics} \textbf {103} 661$-$681 (1981)




\bibitem{Singer} M. van der Put and M. F. Singer,
Differential Galois Theory
Published as Galois Theory of Linear Differential Equations \emph{Grundlehren der mathematischen Wissenschaften} \normalfont \textbf{328}  Springer (2003)




\bibitem{Singer1} M. F. Singer, Introduction to the Galois Theory of Linear Differential Equations

This is an expanded version of the 10 lectures given as the 2006 London Mathematical Society Invited Lecture Series at the Heriot$-$Watt University 31 July$-$4 August 2006












\bibitem{Duval}A. Duval  and M. Loday$-$Richaud,
A propos de l'algotitme de Kovacic
\emph{Tech. Rep.} Universite
de Paris$-$Sud Mathematiques Orsay France
(1989)


\bibitem{Duval1}A. Duval  and M. Loday$-$Richaud,
Kovacic's Algorithm and its Application to Some Families of
Special Functions
\emph{AAECC} \textbf {3} 211$-$246 (1992)


\bibitem{UW} F. Ulmer  and J. A. Weil,  Note on Kovacic's algorithm \emph{J. Symbol. Comput.} \textbf {22}(2)
179$-$200 (1996)


\bibitem{SU1} M. F. Singer  and F. Ulmer, Liouvillian and algebraic solutions of second and third
order linear differential equations \emph{J. Symbol. Comput.} \textbf {16}(1) 37$-$73 (1993)


\bibitem{SU2} M. F. Singer and F. Ulmer,  Necessary conditions for Liouvillian solutions of (third
order) linear differential equations \emph{AAECC} \textbf 6(1) 1$-$22 (1995)



\bibitem{HR} M. van Hoeij, J. F. Ragot, F. Ulmer and J. A. Weil,  Liouvillian solutions of linear
differential equations of order three and higher \emph{J. Symbol Comput.}  \textbf  {28} 589$-$609 (1998)

\bibitem{melas} E. Melas Picard$-$Vessiot$-$Kolchin theory and the CEV model (in preparation)











\end{thebibliography}
\end{document}